\documentclass[%
 reprint,
 amsmath,amssymb,
 aps,
prx,
]{revtex4-2}

\usepackage[utf8]{inputenc}
\usepackage[T1]{fontenc}
\usepackage{amsmath}
\usepackage{mathrsfs}
\usepackage[colorlinks=true, allcolors=blue]{hyperref}

\usepackage{graphicx}
\usepackage{dcolumn}
\usepackage{bm}
\usepackage{svg}

\begin{document}

\preprint{APS/123-QED}

\title{Kinetic route to helicity-constrained decay}

\author{Dion Li}
 \email{dionli@psfc.mit.edu}
\affiliation{Plasma Science and Fusion Center, Massachusetts Institute of Technology, Cambridge, MA 02139, USA}%

\date{\today}

\begin{abstract}
Through two-dimensional, three-velocity-component particle-in-cell simulations of freely decaying subion turbulence, intermittent localized regions with $\mathbf{E} \cdot \mathbf{B} \neq 0$ are found, in the early electron-scale interaction phase, to be statistically associated with decreases in $|H_{V_s}|$, the fixed-gauge structure-integrated magnetic-helicity diagnostic. This structure-level behavior coincides with a decline of the Saffman helicity-variance plateau value $I_H$. Motivated by these observations, we propose a source-compensated, history-dependent helicity density that satisfies an exact local balance identity by construction, enabling Saffman-type two-point correlation integrals, which, under standard flux-decorrelation assumptions, can exhibit intermediate-scale plateaus that are roughly time independent. In the simulations, such plateaus are observed to remain approximately invariant over the measured kinetic interval even as $I_H$ evolves during the early kinetic stage. Under approximate single-scale self-similarity, the plateau behavior of the magnetic integral is consistent with the two-dimensional decay constraint $BL \sim \text{const}$. For initially net-helical configurations, we observe rapid development of mixed-signed magnetic-helicity patches and a decrease of the global fractional helicity, such that the decay over the kinetic interval is again most consistent with the cancellation-dominated scaling constraint.
\end{abstract}

\maketitle

\section{Introduction}\label{sec:introduction}
In ideal magnetohydrodynamics (MHD), the total magnetic helicity $H_V = \int_V dV \; h$, with $h = \mathbf{A} \cdot \mathbf{B}$ the magnetic-helicity density, is gauge invariant and conserved in a simply connected domain when $\partial V$ is a magnetic, impermeable boundary, or otherwise arranged so boundary terms vanish~\citep{Elsasser1956,Woltjer1958_a,Berger1984,Boozer1986,Arnold2021}. As a global topological measure of field-line twist, writhe, and linkage~\citep{Moffatt1969,Moffatt1992}, $H_V$ plays a central organizing role in magnetic relaxation and turbulent decay. In high-Lundquist-number plasmas, magnetic helicity often remains comparatively well conserved, provided boundary fluxes are weak~\citep{Taylor1974,Taylor1986,Ji1995,Guo2004}. Because resistive dissipation is gradient weighted and thus acts most strongly at small scales~\citep{Blackman2016}, and because helical turbulence tends to transfer magnetic helicity to larger scales~\citep{Frisch1975,Biskamp1999_a,Mller2000,Christensson2001,Alexakis2006,Pouquet2019}, the total magnetic helicity usually decays more slowly than magnetic energy at high Lundquist numbers, making it a natural approximate invariant that constrains the late-time decay of magnetic turbulence~\citep{Matthaeus1980,Stribling1995,Banerjee2004,Durrer2013,Subramanian2016}.

For fully helical fields dominated by a single energy-containing scale, $L$ (see, e.g., Eq.~\ref{eq:intscale}), conservation of magnetic helicity implies $B^2 L \sim \text{const}$~\citep{Field2000,Christensson2001,Caprini2014}, with $B$ (or $B_{\text{rms}}$) the rms magnetic field. A distinct but common situation arises when the configuration is globally nonhelical in the sense that net magnetic helicity cancels, even though strong local magnetic-helicity fluctuations persist. For such configurations, \citet{Hosking2021} proposed a Saffman-type integral built from two-point correlations of magnetic-helicity density,
\begin{align}
\mathcal{I}_H (R) \equiv \int_{V_R} d^3 r \; \langle h (\mathbf{x}) h (\mathbf{x} + \mathbf{r}) \rangle,
\label{eq:IHdef}
\end{align}
with $\langle \cdot \rangle$ an ensemble average. Under suitable scale separation, $\mathcal{I}_H (R)$ can develop an intermediate-range plateau that is approximately conserved. Indeed, \citet{Zhou2022} confirmed that $\mathcal{I}_H (R)$ approaches an $R$-independent asymptote $I_H$ for $L \ll R \ll L_{\text{sys}}$~\citep[see also][]{Brandenburg2025,Hew2026}, with $L_{\text{sys}}$ the system domain size. Under the same scale-separation/localization assumptions, conservation of $I_H$ yields $B^4 L^5 \sim \text{const}$ in three dimensions (3D)~\citep{Hosking2021,Brandenburg2023,Armua2023,Hosking2023}. 

A key limitation of these helicity-based constraints is that many astrophysical and space plasmas do not remain in the ideal-MHD regime across all dynamically active scales~\citep{Yamada2010,Bruno2013,Alexandrova2013,Verscharen2019,Sahraoui2020}. Once the ideal-MHD ordering breaks down, magnetic helicity need not be conserved even in the absence of explicit resistive dissipation, because the magnetic-helicity density obeys the general evolution law~\citep{Berger1984,Finn1985}:
\begin{align}
\partial_t h - c \nabla \cdot (\mathbf{A} \times \mathbf{E} - \varphi \mathbf{B}) = - 2c \mathbf{E} \cdot \mathbf{B},
\label{eq:Hevoleqn}
\end{align}
with $\varphi$ the electric potential. The source term $-2c \mathbf{E} \cdot \mathbf{B}$ implies that any localized nonideal region with $\mathbf{E} \cdot \mathbf{B} \neq 0$ permits magnetic-helicity change~\citep{Schindler1988}. In resistive MHD, this departure from ideality is associated with magnetic diffusion and is often characterized by the Lundquist number $S_L \sim L v_A/\eta$, with $\eta$ the magnetic diffusivity~\citep{Ji1998,Priest2000,Biskamp2000}. In the collisionless subion regime considered here, in contrast, there is no unique scalar magnetic diffusivity and hence no single Lundquist number analogous to resistive MHD~\citep{Vasyliunas1975,Biskamp1997,Shay2001,Khotyaintsev2019,Liu2025_a}. Rather, electron frozen-in breaking is localized in electron-scale nonideal regions, including electron diffusion regions (EDRs)~\citep{Zweibel2009,Yamada2010}, and is supported by pressure-tensor and inertial effects in the generalized Ohm's law (see, e.g., Eq.~\ref{eq:Ohm}).

Such localized nonideal regions arise naturally in kinetic turbulence. A broad body of kinetic~\citep{Wan2012,Wan2015,TenBarge2013,Haggerty2017} and hybrid-kinetic~\citep{Servidio2012,Servidio2014,Franci2015,Franci2018,SorrisoValvo2018} work, together with \textit{in situ} space-plasma evidence~\citep{Osman2012,Osman2014,Chasapis2015,Chasapis2018,Vrs2017,Bandyopadhyay2020}, has shown in a wide variety of settings that turbulence rapidly generates intermittent current sheets at ion scales, many of these sheets reconnect, and reconnection commonly coincides with localized dissipation and heating signatures. In sufficiently thin current sheets, reconnection can occur in an ``electron-only'' mode~\citep{Zweibel2009,Yamada2010}, with observational~\citep{Yordanova2016,Phan2018,Stawarz2019} and numerical~\citep{SharmaPyakurel2019,Vega2020,Mallet2020,Califano2020,Liu2025_b} evidence for electron-only reconnection having grown substantially over the last decade. Thus, the classical magnetic-helicity constraint that organizes MHD decay competes with localized kinetic nonideality at the very scales where turbulence becomes most intermittent. This competition motivates two closely related questions: (1) How does collisionless, reconnection-mediated nonideality modify magnetic helicity at subion scales? (2) If the traditional ideal-MHD magnetic-helicity constraint is compromised, is there a first-principles kinetic analog that can still provide a practically useful decay constraint?

In this paper, we provide possible answers to both questions in the specific setting of freely decaying two-dimensional, three-velocity-component (2D3V) turbulence at subion scales, with all three components of the electromagnetic fields retained. In Sec.~\ref{sec:Hevol}, using fixed-gauge structure diagnostics, we find evidence that during the early kinetic phase, intermittent regions with $\mathbf{E} \cdot \mathbf{B} \neq 0$ are statistically associated with structure-handedness proxies and with decreases in $|H_{V_s}|$, the structure-integrated magnetic-helicity diagnostic. We further show that this early structure-level depletion coincides with a decline of the usual Saffman helicity-variance plateau value $I_H$, consistent with a reduction of magnetic-helicity-density variance. We do not interpret $H_{V_s}$ as a gauge-independent subvolume helicity invariant; rather, it is used as a diagnostic of structure-level handedness and helicity change in the chosen gauge, while the broader helicity-evolution statements are based on spectral fractional helicity and finite-window measures interpreted under the stated localization assumptions.

Motivated by the behavior observed in Sec.~\ref{sec:Hevol}, we then introduce in Sec.~\ref{sec:canonicalhelicity} a fully kinetic, source-compensated, history-dependent helicity density that satisfies an exact local balance identity by construction. Although this quantity is an accounting reformulation rather than a topological invariant, it motivates kinetic Saffman-type two-point integrals that can exhibit approximately time-independent intermediate-scale plateaus even while the usual helicity-based plateau value $I_H$ evolves. When referenced to a specific interaction interval, this construction also admits an event-local helicity-balance interpretation; over longer intervals, however, its meaning is cumulative and is assessed through the observed plateau behavior of the corresponding correlation integral.

We focus primarily on the magnetic component of the conserved density because it reduces to the usual magnetic-helicity density in the ideal-MHD limit and is most directly relevant when flows are initially absent or subdominant. Under approximate single-scale self-similarity, the measured plateau behavior is consistent with the two-dimensional (2D) decay constraint $BL \sim \text{const}$, and we present numerical evidence compatible with this scaling. We examine both initially nonhelical and initially net-helical configurations, corresponding to $\sigma_0 = 0$ and $1$, in 2D3V particle-in-cell (PIC) simulations performed at both reduced ($m_i/m_e = 25$) and realistic ($m_i/m_e \approx 1836$) mass ratios.

For the initially net-helical case, discussed in Sec.~\ref{sec:nethelical}, we observe that turbulent reconnection quickly develops mixed-signed magnetic-helicity patches. As a result, the kinetic-scale dynamics tend toward an effectively nonhelical state, and the measured decay is again most consistent with $BL \sim \text{const}$.

More broadly, the aim of this work is not only to characterize one freely decaying 2D3V PIC system, but to identify a kinetic route by which helicity-constrained decay may be reformulated when ideal-MHD conservation is compromised by localized nonideality. In that sense, the source-compensated construction developed here is intended as a general diagnostic and interpretive framework that can be tested in other kinetic and weakly collisional plasma settings, including more realistic 3D turbulence, where subion reconnection and intermittent dissipation are also expected to be important.

\section{Magnetic-helicity evolution at subion scales}\label{sec:Hevol}
To separate the physical argument from the later formal construction, we begin with structure-level diagnostics of subion nonideality and helicity change, then connect those diagnostics to the source-compensated formulation introduced in Sec.~\ref{sec:canonicalhelicity}.

\subsection{Structure-level diagnostics}
To assess the implications of subion nonideality on magnetic helicity, we analyze our freely decaying 2D3V PIC simulations with $\partial_z = 0$; the results below span both initially nonhelical and net-helical configurations and both reduced and realistic ion-electron mass ratios (see Appendix~\ref{appA} for the precise definition of $\sigma$ and the full numerical setup). The correlations reported below should therefore be interpreted as tendencies within this geometry and initial-condition class. In this geometry, it is convenient to write~\citep{Strauss1976}
\begin{align}
\mathbf{B} = \nabla A_z \times \hat{\mathbf{z}} + \hat{\mathbf{z}} B_z, \label{eq:Azdef}
\end{align}
with $A_z (x,y)$ the out-of-plane component of the magnetic vector potential. Equation~\ref{eq:Azdef} implies $\mathbf{B} \cdot \nabla A_z = 0$, so the level sets $A_z = \text{const}$ are magnetic surfaces. This makes $A_z$-contour-bounded structures a natural diagnostic object in 2D3V. Hence, we define the boundary of an individual coherent structure $s$ at each time by a closed $A_z$ contour and consider the fixed-gauge diagnostic~\footnote{For the diagnostics used here, $\mathbf{A}$ is reconstructed in periodic Coulomb gauge from $\mathbf{B}$; details are given in Appendix~\ref{appA}.}
\begin{align}
H_{V_s} = \int_{V_s} dV \; \mathbf{A} \cdot \mathbf{B}. \label{eq:HVs}
\end{align}
Because $V_s$ is defined by an instantaneous $A_z$ contour, it is time dependent and $dH_{V_s}/dt$ contains, in general (see, e.g., Eq.~\ref{eq:Hevoleqn}), (1) the nonideal volume contribution $-2c \int_{V_s} dV \; \mathbf{E} \cdot \mathbf{B}$ and (2) transport/exchange contributions across the moving structure boundary. The $A_z = \text{const}$ choice implies $\mathbf{B} \cdot \hat{\mathbf{n}} = 0$ on the structure sidewall, which removes the $\varphi \mathbf{B}$ contribution there, but it does not in general eliminate $\mathbf{A} \times \mathbf{E}$ transport or the additional moving-boundary term associated with $V_s = V_s(t)$. In what follows, we use $-2c \int_{V_s} dV \; \mathbf{E} \cdot \mathbf{B}$ as the primary source term for magnetic-helicity change within interacting structures and interpret the remaining contributions as magnetic-helicity redistribution among neighboring regions.

\subsection{Nonideal proxy and magnetic energy conversion}
In the context of collisionless reconnection, the generalized Ohm's law gives the nonideal electric field in the electron frame~\citep{Zweibel2009,Yamada2010,Treumann2013}:
\begin{align}
\mathbf{E}^{\prime} = \mathbf{E} + \frac{\mathbf{u}_e \times \mathbf{B}}{c} = - \frac{1}{en} \nabla \cdot \pmb{\Pi}_e \nonumber \\
- \frac{m_e}{e} [ \partial_t \mathbf{u}_e + (\mathbf{u}_e \cdot \nabla) \mathbf{u}_e ], \label{eq:Ohm}
\end{align}
supported in the EDR by the divergence of the electron pressure tensor $\pmb{\Pi}_e$ and by electron inertia~\footnote{In the numerical diagnostics, we evaluate $\mathbf{E}^{\prime}$ directly from the simulation fields as $\mathbf{E}^{\prime} = \mathbf{E} + \mathbf{u}_e \times \mathbf{B}/c$; Eq.~\ref{eq:Ohm} is used to identify the kinetic terms supporting $\mathbf{E}^{\prime}$, not as the numerical prescription for computing it.}. Here, $e > 0$ and $\mathbf{u}_e$ is the electron flow. In our simulations, during the interactions of coherent structures at subion scales and near $d_e = (m_e c^2/4\pi n e^2)^{1/2}$, the electron skin depth or inertial length, regions identified operationally as EDR-like exhibit a statistical association between the sign of $\mathbf{E} \cdot \mathbf{B}$ and structure handedness as measured by a current-helicity proxy~\citep{Seehafer1990,Hagino2004,Pevtsov2014}, e.g., $\mathbf{J} \cdot \mathbf{B}$ or~\citep[e.g.,][]{Woltjer1958_a,Taylor1986,Wiegelmann2012}
\begin{align}
\alpha \equiv \mathbf{J} \cdot \mathbf{B}/|\mathbf{B}|^2,
\end{align}
with $\mathbf{J}$ the current density from species flows. 

Decomposing $\mathbf{J}$ into components parallel and perpendicular to $\hat{\mathbf{b}} \equiv \mathbf{B}/|\mathbf{B}|$ gives
\begin{align}
\mathbf{J} = (\mathbf{J} \cdot \hat{\mathbf{b}}) \hat{\mathbf{b}} + (\mathbb{I} - \hat{\mathbf{b}} \otimes \hat{\mathbf{b}}) \cdot \mathbf{J} = J_{\parallel} \hat{\mathbf{b}} + \mathbf{J}_{\perp},
\end{align}
so that $J_{\parallel} = \alpha |\mathbf{B}|$ and $\alpha \mathbf{E} \cdot \mathbf{B} = E^{\prime}_{\parallel} J_{\parallel}$. 

\begin{figure*}
\begin{center}
\includegraphics[width=1.0\textwidth]{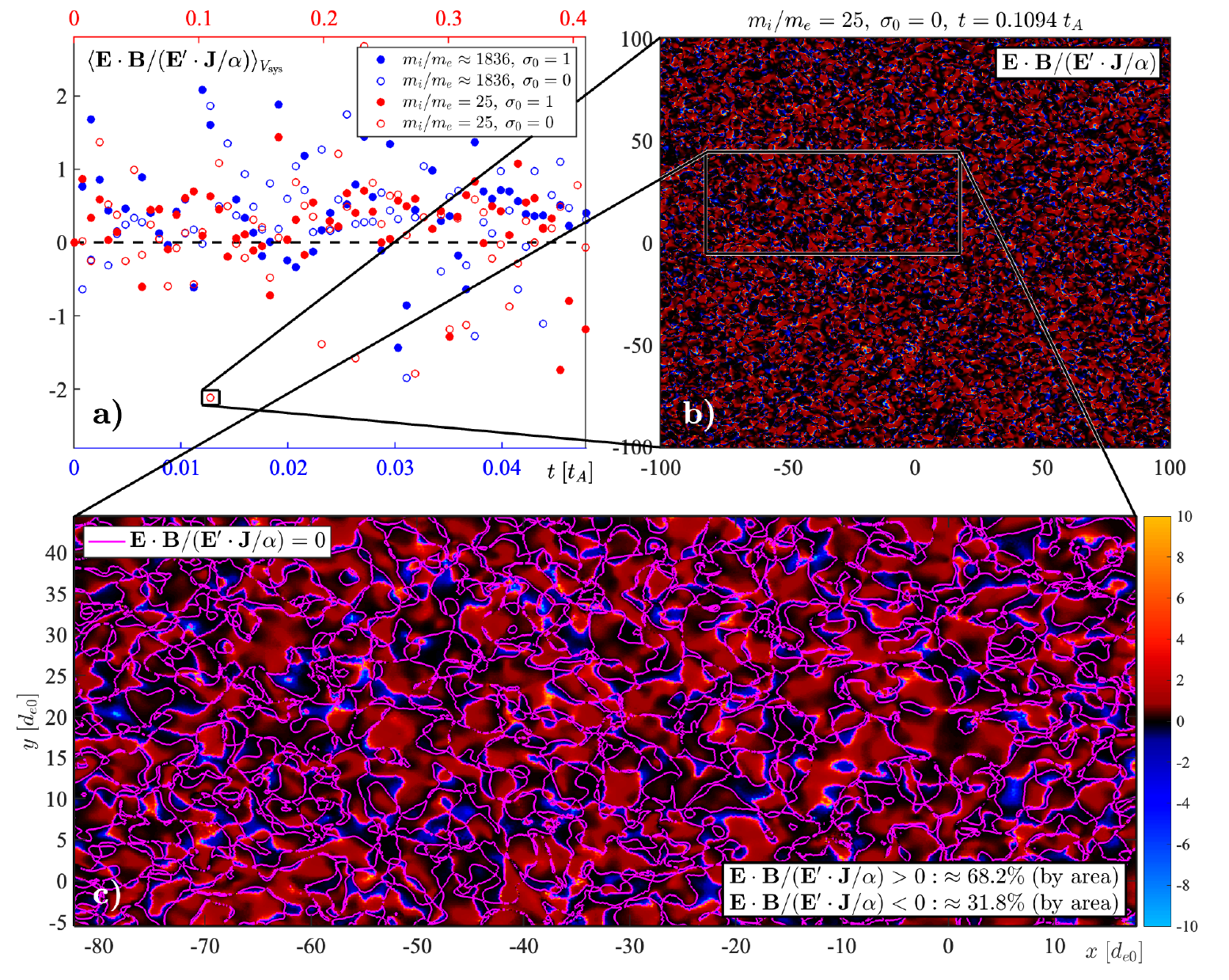}
\caption{\label{fig:EJsign} Numerical evidence of Eqs.~\ref{eq:EJsign_scaling} and~\ref{eq:sign} in all simulations presented in this work (see Table~\ref{tab:table1}). Panel (a) plots the domain average $\langle \mathbf{E} \cdot \mathbf{B}/(\mathbf{E}^{\prime} \cdot \mathbf{J}/\alpha) \rangle_{V_{\text{sys}}}$ at times spaced by $\Delta t \approx [6.84 \times 10^{-3}, 7.98 \times 10^{-4}] \; t_A$ for $m_i/m_e = 25$ and $\approx 1836$, respectively. Figure~\ref{fig:EJsign_hist_app} shows the corresponding distributions of these panel (a) values over sampled output times for each run. Included as the black dashed line is when $\langle \mathbf{E} \cdot \mathbf{B}/(\mathbf{E}^{\prime} \cdot \mathbf{J}/\alpha) \rangle_{V_{\text{sys}}} = 0$. Panels (b) and (c) show the strongest outlier from panel (a), first over the full domain and then over the subdomain with the largest negative average value. Even in this case, the majority of the subdomain satisfies Eq.~\ref{eq:sign} by area. Here, $d_{e0} = d_e (n_0) = (m_e c^2/4\pi n_0 e^2)^{1/2}$, and $n_0$ is the initial, uniform ion and electron number density.}
\end{center}
\end{figure*}

A widely used electron-frame energy-conversion proxy for identifying and characterizing EDRs in collisionless reconnection is the quantity $\mathbf{E}^{\prime} \cdot \mathbf{J}$~\citep{Zenitani2011_a,Zenitani2011_b,Zenitani2012,Burch2016,Torbert2018,Genestreti2025}. Although the relative importance of its parallel and perpendicular contributions depends on reconnection geometry (antiparallel/low-guide cases tend to be perpendicular dominated near the inner EDR~
\citep{Yamada2014,Wilder2018,Fox2018}, whereas guide-field reconnection can be parallel dominated~\citep{Dahlin2014,Dahlin2016,Eriksson2016,Eastwood2018}), we find empirically in our $\beta \equiv 8\pi (n_i T_i + n_e T_e)/B^2 \sim 1$ simulations that
\begin{align}
\mathbf{E} \cdot \mathbf{B} \sim \mathbf{E}^{\prime} \cdot \mathbf{J}/\alpha \label{eq:EJsign_scaling}
\end{align}
in EDR-like regions, consistent with cases where the parallel conversion term contributes substantially to $\mathbf{E}^{\prime} \cdot \mathbf{J}$, and that the signs predominantly agree,
\begin{align}
\text{sgn} (\mathbf{E} \cdot \mathbf{B}) \simeq \text{sgn} (\mathbf{E}^{\prime} \cdot \mathbf{J}/\alpha), \label{eq:sign}
\end{align}
with intermittent exceptions, demonstrated in Fig.~\ref{fig:EJsign}. Figure~\ref{fig:EJsign}(a) shows the time evolution of the domain-averaged ratio
\begin{align}
Q(t) \equiv \langle \mathbf{E} \cdot \mathbf{B}/(\mathbf{E}^{\prime} \cdot \mathbf{J}/\alpha) \rangle_{V_{\text{sys}}}, \label{eq:ratio}
\end{align}
while Fig.~\ref{fig:EJsign_hist_app} shows the corresponding distributions of $Q(t)$ over sampled output times for each simulation run. Across all four runs, the sampled values remain concentrated within an order-unity range and are predominantly positive, although intermittent negative domain averages do occur. Figure~\ref{fig:EJsign}(b, c) also illustrates an important limitation of using the unweighted domain average of the ratio
$\mathbf{E} \cdot \mathbf{B}/(\mathbf{E}^{\prime} \cdot \mathbf{J}/\alpha)$
as a sign-alignment diagnostic. In the strongest negative outlier, Eq.~\ref{eq:sign} is still satisfied over a majority of the selected subdomain by area, showing that same-sign behavior remains spatially prevalent. However, the negative value of the unweighted average indicates that a smaller number of intense antialigned patches, or patches in which the denominator is small, can dominate the arithmetic mean. We therefore use the area fraction only to diagnose the spatial prevalence of the pointwise sign relation, not to claim dominance of the corresponding energetic or helicity contribution. The negative outliers are retained as evidence of intermittent departures from Eq.~\ref{eq:sign}, which may include non-reconnection-associated contributions or regions outside the simple EDR-like interpretation.

We relate $\mathbf{E}^{\prime} \cdot \mathbf{J}$ to magnetic energy conversion through Poynting's theorem~\citep{Poynting1884}
\begin{align}
\partial_t \bigg( \frac{E^2 + B^2}{8\pi} \bigg) + \frac{c}{4\pi} \nabla \cdot (\mathbf{E} \times \mathbf{B}) = - \mathbf{E} \cdot \mathbf{J}, \label{eq:poynting}
\end{align}
and in our simulations we find $E_{\text{rms}}^2/B_{\text{rms}}^2 \le 1$ throughout (see Fig.~\ref{fig:EvsB_app}). Thus, the electric-field energy is not dominant in the field-energy budget, and $\mathbf{E}^{\prime} \cdot \mathbf{J}$ provides a useful electron-frame proxy for magnetic energy conversion at the order-of-magnitude level used here.

Over a selected interaction area, the exact electromagnetic field-energy exchange term from Poynting's theorem is $- \int_{V_s} dV \; \mathbf{E} \cdot \mathbf{J}$. In the reconnection/EDR interpretation, however, we use $\int_{V_s} dV \; \mathbf{E}^{\prime} \cdot \mathbf{J}$ as the corresponding electron-frame energy-conversion proxy. These two quantities coincide to leading order when the ion response is negligible near electron scales, since $\mathbf{J} \simeq -en \mathbf{u}_e$~\citep{Zweibel2009,Phan2018,SharmaPyakurel2019} implies $(\mathbf{u}_e \times \mathbf{B}) \cdot \mathbf{J} \simeq 0$ and hence $\mathbf{E}^{\prime} \cdot \mathbf{J} \simeq \mathbf{E} \cdot \mathbf{J}$. As with the transport/exchange contributions to $H_{V_s}$, the remaining terms are interpreted as field-energy redistribution among neighboring regions. 

Together with these approximations, Eq.~\ref{eq:poynting} then implies
\begin{align}
\langle \mathbf{E}^{\prime} \cdot \mathbf{J} \rangle_{V_s} \sim - \langle \dot{E}_B \rangle_{V_s} > 0, \label{eq:EBapprox}
\end{align}
where $\langle \cdot \rangle_{V_s}$ denotes a spatial average over $V_s$ and $E_B \equiv B^2/8\pi$ is the magnetic-energy density. The inequality in Eq.~\ref{eq:EBapprox} expresses the decaying-turbulence regime, i.e., $\dot{E}_B < 0$ (see, e.g., the $B_{\text{rms}}^4$ curve in Fig.~\ref{fig:Saffman_evol_theory}).

\subsection{Sign alignment, helicity depletion, and implications for $I_H$}
We now connect the nonideal proxy to structure-level helicity diagnostics and to the evolution of the magnetic-helicity variance. From Eqs.~\ref{eq:EJsign_scaling},~\ref{eq:sign}, and~\ref{eq:EBapprox}, we find that the structure-averaged nonideality satisfies the order-of-magnitude tendency
\begin{align}
\langle \mathbf{E} \cdot \mathbf{B} \rangle_{V_s} \sim - \alpha_s^{-1} \langle \dot{E}_B \rangle_{V_s}, \label{eq:dotEBkeff}
\end{align}
provided that $\alpha$ is approximately sign coherent within the selected $A_z$-bounded structures. Here, 
\begin{align}
\alpha_s \equiv H_{C_s} \bigg( \int_{V_s} dV \; B^2 \bigg)^{-1},
\end{align}
where
\begin{align}
H_{C_s} = \int_{V_s} dV \; \mathbf{J} \cdot \mathbf{B} \label{eq:HCs}
\end{align}
is the current helicity in $V_s$. To quantify the handedness of each magnetic structure in a way that is directly comparable to $\alpha_s$, we compute for every $A_z$-bounded structure the structure-integrated proxies $H_{V_s}$ and $H_{C_s}$ and define a sign-alignment diagnostic 
\begin{align}
S \equiv \text{sgn}(H_{V_s}) \text{sgn}(H_{C_s}) \in \{ \pm 1 \}. \label{eq:contour_sign}
\end{align}

\begin{figure}[b]
\includegraphics[width=0.5\textwidth]{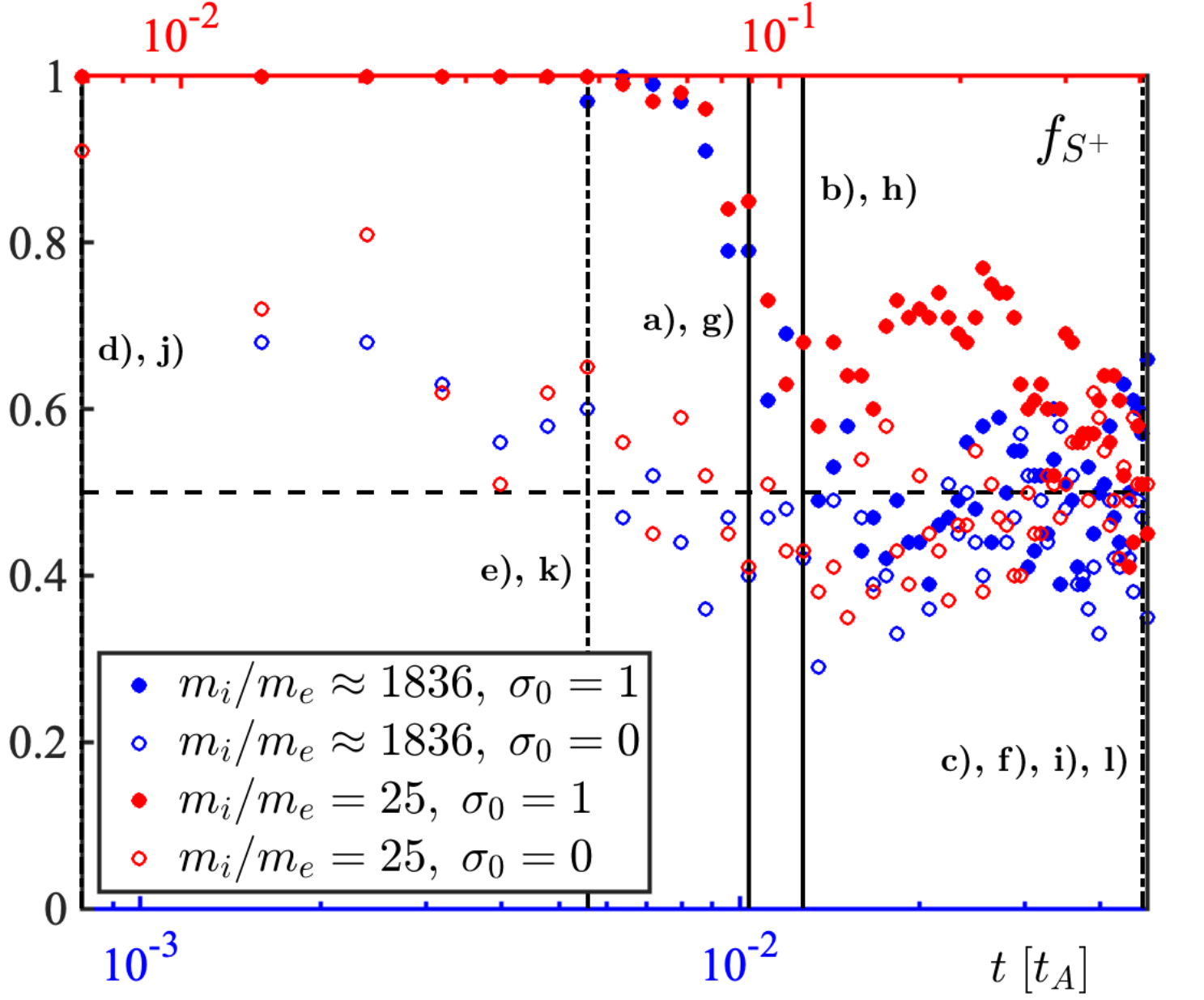}
\caption{\label{fig:contour_align_plot} Evolution of $f_{S^+}$, the fraction of structures bounded by $A_z$ contours for which $S \equiv \text{sgn}(H_{V_s}) \text{sgn}(H_{C_s}) = + 1$, for all 2D3V PIC simulations presented in this work. Indicated by the solid and dash-dotted black lines for initially net-helical and nonhelical configurations, respectively, are the times plotted in Fig.~\ref{fig:contour_align}, with annotations indicating the corresponding subplot letter. The black dashed line indicates $f_{S^+} = 0.5$. }
\end{figure}

\begin{figure*}
\begin{center}
\includegraphics[width=1.0\textwidth]{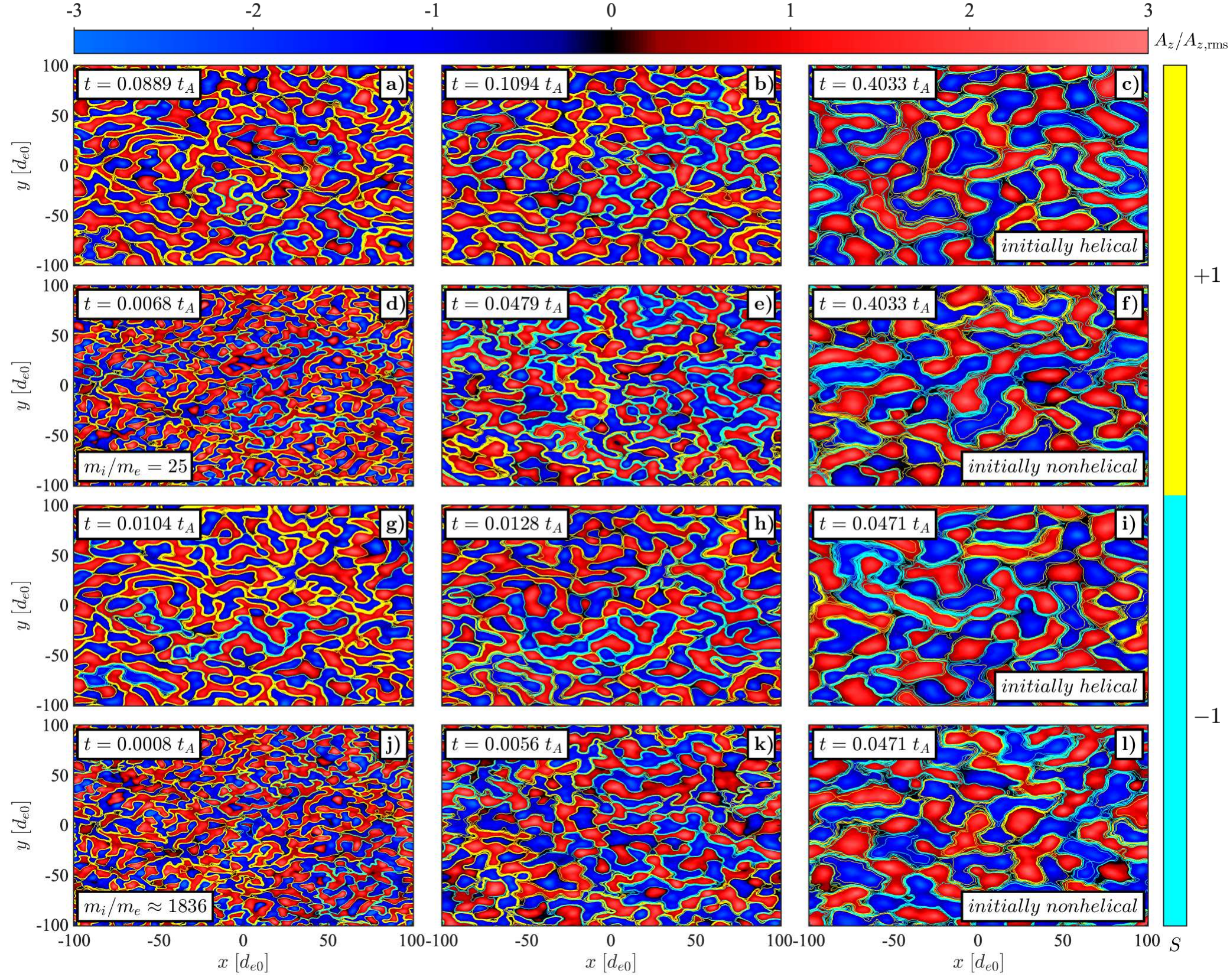}
\caption{\label{fig:contour_align} Representation of the structure-by-structure helicity sign alignment in our 2D3V PIC simulations. Each panel shows the out-of-plane flux function $A_z (x,y)$ (red-blue colormap) together with closed $A_z = \text{const}$ contours (flux surfaces) that define individual coherent magnetic structures. For each structure $s$, we compute the structure-integrated helicity proxies Eqs.~\ref{eq:HVs} and~\ref{eq:HCs}, and color the structure boundary by the sign alignment (Eq.~\ref{eq:contour_sign}; yellow denotes aligned signs, $S = +1$, and cyan denotes antialigned signs, $S = -1$). Shown are three representative times (early (left column)/intermediate (middle column)/late (right column), indicated by the solid and dash-dotted black lines in Fig.~\ref{fig:contour_align_plot}). At early times—when subion nonideality and reconnection-driven interactions are strongest—structure boundaries are predominantly yellow, indicating that a large majority of coherent structures satisfy Eq.~\ref{eq:sign_alignment}. At later times, the population becomes mixed, showing comparable yellow/cyan, and in some cases majority cyan. }
\end{center}
\end{figure*}

Our structure-by-structure measurements show that, at early times when subion nonideality is strongest, a large majority of $A_z$-bounded structures exhibit the same handedness in magnetic- and current-helicity proxies, i.e., 
\begin{align}
\text{sgn}(H_{V_s}) \simeq \text{sgn}(H_{C_s}), \label{eq:sign_alignment}
\end{align}
as reported by Fig.~\ref{fig:contour_align_plot}, plotting the evolution of $f_{S^+}$, the fraction of structures bounded by $A_z$ contours for which $S = + 1$, and Fig.~\ref{fig:contour_align}, showing colormap representations of the structure-by-structure helicity sign alignment. At later times, occurring earlier for initially nonhelical configurations, the alignment becomes mixed, indicating that the ensemble no longer maintains a strong one-to-one sign correspondence between $\mathbf{A} \cdot \mathbf{B}$ and $\mathbf{J} \cdot \mathbf{B}$ at the structure level. 

A natural interpretation of the observed early-time structure-level sign alignment is that, during the strongly nonideal phase, the field-aligned, reconnection-relevant portion of the current remains largely organized by the structures' handedness. Under isotropic conditions, this is broadly consistent with the Amp\`ere-current expectation that current helicity is a $k^2$-weighted version of magnetic helicity~\citep{Brandenburg2001}. Equation~\ref{eq:contour_sign} is therefore not intended as a scale-independent description about the full turbulent spectrum. Rather, sign agreement between $H_{V_s}$ and current-helicity proxies is expected only when the selected structure is dominated by a relatively narrow, effectively single-signed band of $H_k$. If substantial power is distributed across widely separated scales, or if $H_k$ changes sign across the excited band, then the current-helicity proxy weights the smaller scales more strongly, and the alignment can degrade or even reverse. Since our diagnostic uses the physical current $\mathbf{J} = \sum_{\iota} q_{\iota} n_{\iota} \mathbf{u}_{\iota}$~\footnote{In our PIC simulations $\mathbf{u}_i$ and $\mathbf{u}_e$ are outputted as proper velocities, i.e., $\mathbf{u} = \gamma \mathbf{v}$, with $\gamma = (1 - v^2/c^2)^{-1/2}$ the Lorentz factor. Since the flows we observe in our numerical experiments are only mildly relativistic, with $\gamma \simeq 1$ over the domains of interest, replacing $\mathbf{v}$ with the proper velocity in the moment-based current introduces only $\mathcal{O} (\gamma - 1)$ corrections and does not significantly affect the sign-based helicity-alignment diagnostics.}, with $\iota = i, e$ the particle species index, rather than the Amp\`ere current, this interpretation should be regarded as suggestive rather than exact~\footnote{Strictly, Maxwell-Amp\`ere implies that the difference between the Amp\`ere current $\mathbf{J}_A = (c/4\pi) \nabla \times \mathbf{B}$ and the physical particle current is the displacement-current contribution, so the $k^2 H_k$ argument applies directly to $\mathbf{J}_A \cdot \mathbf{B}$ and only qualitatively to the physical-current diagnostic used here.}. We therefore interpret Eq.~\ref{eq:sign_alignment} only as an early-time, interaction-scale tendency for the $A_z$-bounded structures studied here, not as a universal large-scale diagnostic. Consistent with this interpretation, the alignment becomes mixed at later times, as shown in Figs.~\ref{fig:contour_align_plot} and~\ref{fig:contour_align}.

Combining Eqs.~\ref{eq:Hevoleqn},~\ref{eq:dotEBkeff}, and~\ref{eq:sign_alignment}, we arrive at
\begin{align}
|H_{V_s}|^{-1} \frac{d}{dt} \bigg( \frac{1}{2} H_{V_s}^2 \bigg) \sim 8\pi L \int_{V_s} dV \; \dot{E}_B < 0 \label{eq:Hscalingeqn}
\end{align}
during the early kinetic phase, since
\begin{align}
|\alpha|^{-1} \sim (4\pi/c) |\mathbf{B} \cdot (\nabla \times \mathbf{B})|^{-1} |\mathbf{B}|^2 \sim 4\pi L/c.
\end{align}
We interpret Eq.~\ref{eq:Hscalingeqn} as indicative that magnetic-helicity depletion is not merely a consequence of global cancellation, but a dynamically preferred outcome correlated with magnetic energy conversion within the interaction area. In this sense, early-time reconnection at subion scales acts as a mechanism that reduces the helical content of coherent structures while magnetic energy is simultaneously redistributed and dissipated. Because structure boundaries are time dependent, magnetic helicity can also be exchanged between neighboring regions by transport; nevertheless, Eq.~\ref{eq:Hscalingeqn} suggests that, during the dominant interaction intervals, the net effect of nonideality is to bias the evolution toward $|H_{V_s}|$ depletion. 

\begin{figure*}
\begin{center}
\includegraphics[width=1.0\textwidth]{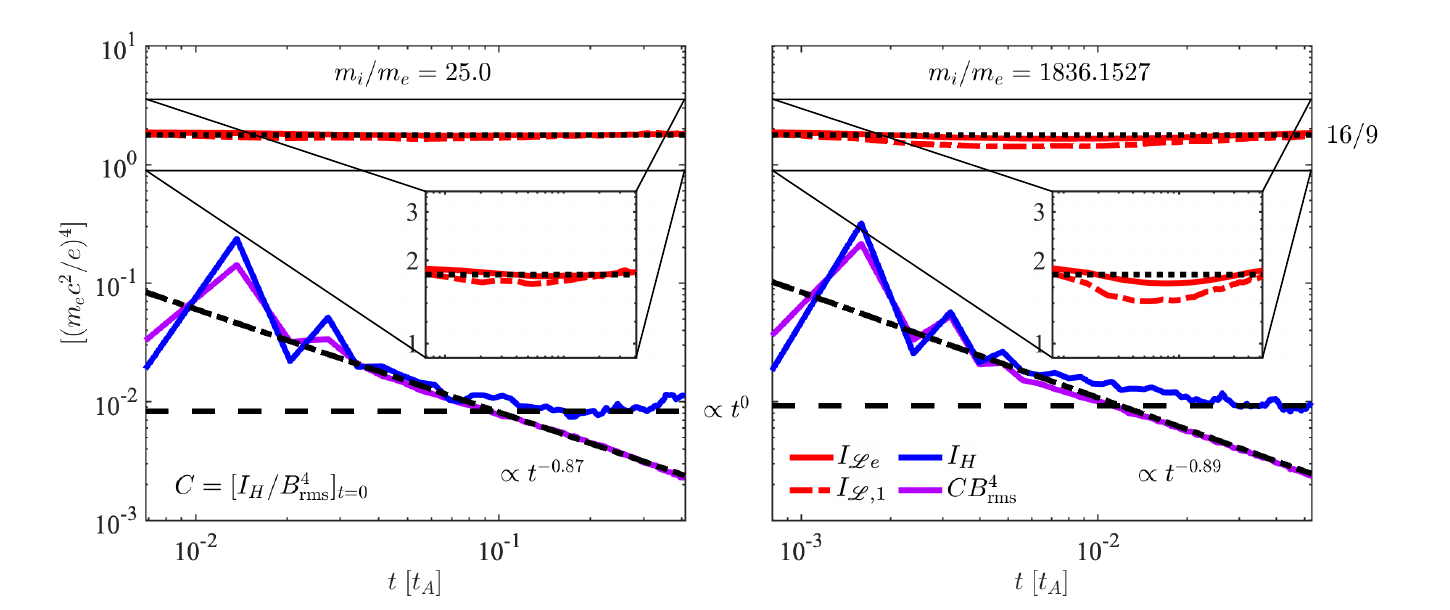}
\caption{\label{fig:Saffman_evol_theory} Time evolution of $I_{\mathscr{L} e}$ (solid red), $I_{\mathscr{L}, 1}$ (dash-dotted red), $I_H$ (solid blue), and $B_{\text{rms}}^4$ (solid purple; multiplied by a proportionality constant equivalent to $2\pi L_h^4 (t = 0)$, indicated by the lower left annotation in the left panel) in globally nonhelical simulations with $m_i/m_e = 25$ (left) and $\approx 1836$ (right). Included (dotted black) is the calculated value of $I_{\mathscr{L}, 1} (t = 0) = I_H (t = 0)$ ($=I_{\mathscr{L} e} (t = 0)$ for initially zero flow), derived in Appendix~\ref{appB}. Also included are the power-law scalings: $\propto t^{-0.87}$ for $m_i/m_e = 25$ and $\propto t^{-0.89}$ for $m_i/m_e \approx 1836$ (dash-dotted black, fitted using the $B_{\text{rms}}^4$ data) and $\propto t^0$ (dashed black). The insets show a close-up of the vertical axis range $\in [1/2, 2] \times 16/9$. }
\end{center}
\end{figure*}

If correlated nonideality systematically reduces $|h|$ within interacting structures, it will tend to reduce magnetic-helicity variance $\langle h^2 \rangle$ and hence $I_H$ (see, e.g., Eq.~\ref{eq:IHdef}). Consistent with this, the most rapid decline of $I_H$ coincides with the regime of high structure-level sign alignment, i.e., predominantly $\text{sgn}(H_{V_s}) = \text{sgn}(H_{C_s})$, whereas $I_H$ becomes approximately stationary once the alignment becomes mixed at later times (see, e.g., Fig.~\ref{fig:Saffman_evol_theory})~\footnote{The interval over which $I_H$ becomes approximately stationary may also be associated with times in which $\mathbf{E} \cdot \mathbf{B}$ becomes comparatively small as the dominant scales move above electron kinetic scales, as discussed in Sec.~\ref{sec:canonicalhelicity}}.

Following the derivation in Appendix~\ref{appB}, the Saffman helicity-variance plateau value may be written approximately as
\begin{align}
I_H \simeq \frac{1}{16} \frac{S_{d - 1}}{(2\pi)^d} \int_0^{\infty} dk \; [1 - \sigma_k^2 (k)] k^{d - 3} F^2 (k). \label{eq:I_H_variancedef}
\end{align}
Here, $d$ is the number of spatial dimensions in the system under consideration, e.g., $d = 2$ for 2D3V. For statistically homogeneous, isotropic, and globally nonhelical fields, this implies
\begin{align}
I_H B_{\text{rms}}^{-4} \propto L_h^{d + 2},
\end{align}
where $L_h$ is the effective magnetic-helicity-density correlation length, defined in Appendix~\ref{appB}. We argue that subion nonideality impedes the transfer of helical content into larger structures, so that magnetic-helicity-density fluctuations may remain controlled by a kinetic correlation length $L_h$ rather than by the growing outer scale $L$. Consistent with this picture, we find for our globally nonhelical initial conditions that $L_h \simeq \text{const}$ as indicated in Fig.~\ref{fig:Saffman_evol_theory}, implying $I_H \propto B_{\text{rms}}^4$ at early times, in the same regime where $H_{V_s}$ and $H_{C_s}$ are sign aligned. Also shown in Fig.~\ref{fig:Saffman_evol_theory} are the approximately conserved quantities $I_{\mathscr{L} e}$ and $I_{\mathscr{L}, 1}$ that we introduce in Sec.~\ref{sec:canonicalhelicity}.

Throughout this section, we focused on the early, strongly nonideal phase of our 2D3V decaying turbulence simulations, during which intermittency and electron-scale reconnection signatures emerge rapidly from the initialized field. Because our initial magnetic field is constructed from a finite set of modes concentrated in a narrow band of wave numbers (see Appendix~\ref{appA}), the earliest structures inherit a comparatively narrow distribution of excited scales. This restricted scale content is also why the sign-alignment diagnostic is most informative at early times. Once the turbulence broadens spectrally and mixed-sign magnetic-helicity contributions develop across scales, the correspondence between $H_{V_s}$ and $H_{C_s}$ is expected to weaken. The scalings and sign correlations discussed in this section should therefore be interpreted as empirical tendencies within this class of initial conditions and geometry, and not as universal properties of all subion turbulence regimes, e.g., broadband initial states, driven steady-state turbulence, or fully 3D systems, where boundary fluxes, scale bandwidth, and structure topology can materially alter magnetic-helicity transport and proxy alignment~\citep{Banerjee2016,Alexakis2018,Pouquet2019,Groelj2019,Squire2022,Stawarz2024}.

\section{Source-compensated helicity}\label{sec:canonicalhelicity}

\subsection{Event-local magnetic balance}
A central implication of Sec.~\ref{sec:Hevol} is that early-time subion interactions are characterized by systematic magnetic-helicity depletion, correlated with magnetic energy conversion. Equation~\ref{eq:Hevoleqn} provides a natural bookkeeping identity for quantifying magnetic-helicity change over an interaction interval $t \in [t_0, t_1]$. This motivates introducing
\begin{align}
\mathscr{L}_1 & (\mathbf{x}, t) \equiv h (\mathbf{x}, t) + 2c \nonumber \\
& \times \int_{t_0}^t dt^{\prime} \; \mathbf{E}(\mathbf{x}, t^{\prime}) \cdot \mathbf{B} (\mathbf{x}, t^{\prime}), \label{eq:eventlocal}
\end{align}
which is constant in time in the absence of fluxes and, more generally, obeys a local conservation law when fluxes are retained (see, e.g., Eq.~\ref{eq:hL_l}). Unless otherwise stated, $t_0$ denotes the initial time of the interval over which the corresponding balance or correlation quantity is evaluated.

Because $\mathscr{L}_1$ contains an integral from a reference time $t_0$, it is history dependent. This does not preclude an event-local interpretation: When $t_0$ is chosen at or just before the onset of a specific interaction, the integrated term records the accumulated nonideal contribution of that interaction, so $\mathscr{L}_1$ serves as the relevant helicity-balance variable over that interval. In contrast, when the interval spans many unrelated interactions, $\mathscr{L}_1$ should be interpreted more cumulatively.

For single-scale coherent structures, both the instantaneous $h$ and the accumulated nonideal term naturally have the same macroscopic dimensions as $B^2 L$. The empirical correlation expressed by Eq.~\ref{eq:Hscalingeqn} further suggests that the same interaction-scale length $L$ controlling magnetic energy conversion also controls the accumulation of $- 2c \int_{t_0}^t dt^{\prime} \; \mathbf{E} \cdot \mathbf{B}$ over the event, making it plausible that $\mathscr{L}_1$ provides a quasistructural label whose amplitude is proportional, up to order-unity, event-dependent factors, to $B^2 L$. In this section, we therefore treat 
\begin{align}
\mathscr{L}_1 \sim B^2 L \label{eq:workinghyp}
\end{align}
as a working hypothesis—motivated by the observed early-time coupling between magnetic-helicity depletion and energy conversion—and use it to motivate a Saffman-type correlation integral for deriving scaling constraints in globally nonhelical configurations. This interpretation is intended for intervals in which $t_0$ is chosen relative to the interaction of interest; over longer histories, $\mathscr{L}_1$ remains useful, but its meaning is more cumulative. We discuss, in Appendix~\ref{appC}, the validity of interpreting Eq.~\ref{eq:workinghyp} as indicative that subion interactions between like-helicity structures are $B^2 L$-preserving processes.

\subsection{Fully kinetic source-compensated density}
To rigorously identify an appropriate conserved quantity in the fully kinetic Vlasov-Maxwell system, we revisit the standard ``canonical helicity'' construction~\citep[e.g.,][]{Steinhauer2001,You2012,You2016,Yoon2017,Yoon2025}, in which canonical helicity is defined from the species canonical momentum and thus unifies magnetic and flow topology in a single quantity. In two-fluid/extended MHD regimes—relevant to collisionless reconnection and subion dynamics—the familiar magnetic, cross~\citep{Woltjer1958_b,Frisch1975}, and kinetic~\citep{Moffatt1969} helicity components are generally not separately invariant~\citep{Lingam2015}; rather, they exchange through the model's nonideal couplings and are redistributed by fluxes across scales and boundaries~\citep{Pouquet2022}. Canonical helicity evolution then ``closes'' on these exchanges, so the apparent nonconservation of the individual components is interpreted primarily as internal conversion, plus boundary/nonideal flux terms, rather than a failure of the underlying generalized invariant~\citep{You2012}. Guided by this logic, we seek a fully kinetic analog that reduces to canonical helicity in the appropriate fluid limit and that remains meaningful when fluid notions of velocity and flux freezing become ambiguous. Accordingly, we start from the Vlasov-Maxwell system for each species
\begin{subequations}
\begin{align}
\partial_t f_{\iota} + \mathbf{v} \cdot \nabla f_{\iota} + \frac{q_{\iota}}{m_{\iota}} \bigg( \mathbf{E} + \frac{\mathbf{v}}{c} \times \mathbf{B} \bigg) \cdot \partial_{\mathbf{v}} f_{\iota} = 0, \label{eq:Vlasov}
\end{align}
\begin{align}
c \nabla \times \mathbf{E} = - \partial_t \mathbf{B}, \label{eq:faraday}
\end{align}
\begin{align}
c \nabla \times \mathbf{B} = \partial_t \mathbf{E} + 4\pi \sum_{\iota} q_{\iota} n_{\iota} \mathbf{u}_{\iota}, \label{eq:ampere}
\end{align}
\end{subequations}
with $n_{\iota} = \int d^3 v \; f_{\iota}$ the number density and $\mathbf{u}_{\iota} = n_{\iota}^{-1} \int d^3 v \; \mathbf{v} f_{\iota}$ the bulk flow. Taking the zeroth and first moments of Eq.~\ref{eq:Vlasov} gives
\begin{subequations}
\begin{align}
\partial_t n_{\iota} + \nabla \cdot (n_{\iota} \mathbf{u}_{\iota}) = 0, \label{eq:continuityeq}
\end{align}
\begin{align}
m_{\iota} \partial_t (n_{\iota} \mathbf{u}_{\iota}) & + m_{\iota} \nabla \cdot (n_{\iota} \mathbf{u}_{\iota} \mathbf{u}_{\iota}) \nonumber \\
= q_{\iota} n_{\iota} & \bigg( \mathbf{E} + \frac{\mathbf{u}_{\iota}}{c} \times \mathbf{B} \bigg) - \nabla \cdot \pmb{\Pi}_{\iota}, \label{eq:momeq}
\end{align}
\end{subequations}
which may be manipulated to read
\begin{align}
\partial_t \mathbf{P}_{\iota} = \mathbf{u}_{\iota} \times \pmb{\Omega}_{\iota} -  \frac{c}{q_{\iota}} \bigg( \nabla \mathcal{E}_{\iota} + \frac{1}{n_{\iota}} \nabla \cdot \pmb{\Pi}_{\iota} \bigg), \label{eq:secondmoment}
\end{align}
where $\mathcal{E}_{\iota} \equiv \frac{1}{2} m_{\iota} u_{\iota}^2 + q_{\iota} \varphi$ is the sum of the kinetic and electric energies of a fluid particle of species $\iota$,
$\pmb{\Pi}_{\iota} \equiv m_{\iota} \int d^3 v \; (\mathbf{v} - \mathbf{u}_{\iota}) (\mathbf{v} - \mathbf{u}_{\iota}) f_{\iota}$ is the pressure tensor, and we have defined the renormalized canonical momentum and associated canonical vorticity as, respectively,
\begin{subequations}
\begin{align}
\mathbf{P}_{\iota} \equiv \mathbf{A} + (m_{\iota} c/q_{\iota}) \mathbf{u}_{\iota},
\end{align}
\begin{align}
\pmb{\Omega}_{\iota} \equiv \nabla \times \mathbf{P}_{\iota} = \mathbf{B} + (m_{\iota} c/q_{\iota}) \pmb{\omega}_{\iota},
\end{align}
\end{subequations}
with $\pmb{\omega}_{\iota} \equiv \nabla \times \mathbf{u}_{\iota}$. Note that by starting from Eq.~\ref{eq:Vlasov}, taking velocity moments gives the exact species continuity and momentum equations, i.e., Eqs.~\ref{eq:continuityeq} and~\ref{eq:momeq}, provided that we retain the full pressure tensor as a moment of $f_{\iota}$. Curling Eq.~\ref{eq:secondmoment} gives the canonical vorticity transport
\begin{align}
\partial_t \pmb{\Omega}_{\iota} = \nabla \times (\mathbf{u}_{\iota} \times \pmb{\Omega}_{\iota}) - \frac{c}{q_{\iota}} \nabla \times \bigg( \frac{1}{n_{\iota}} \nabla \cdot \pmb{\Pi}_{\iota} \bigg). \label{eq:vorticity}
\end{align}
Finally, Eqs.~\ref{eq:secondmoment} and~\ref{eq:vorticity} yield the local continuity equation
\begin{align}
\partial_t (\mathbf{P}_{\iota} \cdot \pmb{\Omega}_{\iota}) + \nabla \cdot \mathbf{F}_{\iota} = \pmb{\mathcal{H}}_{\iota} \cdot \pmb{\Omega}_{\iota}, \label{eq:hLeqn1_wsource}
\end{align}
where we have defined the canonical flux,
\begin{align}
\mathbf{F}_{\iota} \equiv \mathbf{P}_{\iota} \times ( \pmb{\mathcal{H}}_{\iota}/2 + c \nabla \varphi ),
\end{align}
with a momentum-balance residual from rearranging the pressure-tensor term via Eqs.~\ref{eq:continuityeq} and~\ref{eq:momeq}:
\begin{align}
\pmb{\mathcal{H}}_{\iota} \equiv (2 c/q_{\iota}) [ \partial_t (m_{\iota} \mathbf{u}_{\iota}) - q_{\iota} \mathbf{E} ].
\end{align}
The volume integral of $\mathbf{P}_{\iota} \cdot \pmb{\Omega}_{\iota}$ from Eq.~\ref{eq:hLeqn1_wsource} is often referred to as the canonical helicity. For two-fluid systems,~\citet{You2012} explicitly frames the ``relative canonical helicity'' as a gauge-invariant quantity when canonical vorticity flux crosses boundaries, and argues that while species helicities can exchange, the total relative canonical helicity is globally invariant in that idealized framework. However, in the fully kinetic Vlasov-Maxwell setting, there is generally no reason for $\mathbf{P}_{\iota} \cdot \pmb{\Omega}_{\iota}$ to be exactly conserved because kinetic effects bundled into $\pmb{\mathcal{H}}_{\iota} \cdot \pmb{\Omega}_{\iota}$ provide sources/sinks, precisely why canonical vorticity flux conservation is conditional in the first place~\citep{You2016,Yoon2025}.

To work around the explicit source term in Eq.~\ref{eq:hLeqn1_wsource}, we introduce a source-compensated density by absorbing the time-integrated source into the field definition, yielding a conservation identity
\begin{align}
\partial_t \mathscr{L}_{\iota} + \nabla \cdot \mathbf{F}_{\iota} = 0, \label{eq:hLeqn1}
\end{align}
with the locally, source-compensated, conserved density
\begin{align}
\mathscr{L}_{\iota} \equiv \mathbf{P}_{\iota} \cdot \pmb{\Omega}_{\iota} - \int_{t_0}^t dt^{\prime} \; (\pmb{\mathcal{H}}_{\iota} \cdot \pmb{\Omega}_{\iota}) (t^{\prime}). \label{eq:h_L_first}
\end{align}
Equation~\ref{eq:hLeqn1} is therefore an exact conservation statement for $\mathscr{L}_{\iota}$ by construction, within the continuum Vlasov-Maxwell equations. This is a mathematically legitimate step, but physically $\mathscr{L}_{\iota}$ is not the usual notion of an invariant because it depends on an arbitrary reference time $t_0$, it is history dependent, i.e., nonlocal in time, and it is explicitly an accounting identity rather than a topological constraint. In the present work, this history dependence has two distinct interpretations: (1) event local when $t_0$ is tied to a particular interaction, and (2) cumulative when $\mathscr{L}_{\iota}$ is used to construct correlation integrals over a broader decay interval. These distinctions matter, particularly when we utilize $\mathscr{L}_{\iota}$ to constrain decay scalings because the canonical helicity literature~\citep[e.g.,][]{Brandenburg2005_a,You2012} treats the source terms as physical channels that genuinely create/destroy or exchange helicities, rather than something to be removed by redefinition. Nevertheless, what is important for our considerations is that Eq.~\ref{eq:hLeqn1} describes an exact conservation law that carries no two-fluid or MHD reduction and holds for fully kinetic Vlasov-Maxwell systems. The need for a source history reflects the absence, in generic unreduced Vlasov-Maxwell dynamics, of a local finite-moment generalized helicity whose instantaneous density removes the kinetic source terms without imposing an additional closure, ordering, or frozen-in structure. Thus $\mathscr{L}_{\iota}$ is best interpreted as a minimal helicity budget for the full kinetic system; its dynamical relevance must then be tested through the Saffman-type plateaus introduced in Sec.~\ref{subsec:IIIC}. 

It is useful to decompose $\mathscr{L}_{\iota}$ into four source-compensated components, $\mathscr{L}_{\iota,l}$ with $l \in \{ 1, 2, 3, 4 \}$, each obeying a continuity equation analogous to Eq.~\ref{eq:hLeqn1},
\begin{align}
\partial_t \mathscr{L}_{\iota, l} + \nabla \cdot \mathbf{F}_{\iota, l} = 0,
\end{align}
with corresponding fluxes $\mathbf{F}_{\iota,l}$ and $\mathscr{L}_{\iota} = \sum_{l=1}^4 \mathscr{L}_{\iota,l}$. The explicit component and flux expressions, their reductions in appropriate fluid limits, and their gauge properties are collected in Appendix~\ref{appD}. For the present paper, the most important component is the magnetic one,
\begin{align}
\mathscr{L}_{\iota, 1} \equiv \mathscr{L}_1 \equiv h + 2c \int_{t_0}^t dt^{\prime} \; (\mathbf{E} \cdot \mathbf{B})(t^{\prime}), \label{subeq:L_1}
\end{align}
whose continuity equation coincides with the magnetic-helicity balance law (Eq.~\ref{eq:Hevoleqn}). In the ideal-MHD limit,
\begin{align}
\mathscr{L}_{1} \underset{\text{ideal MHD}}{\to} h. \label{eq:reduction_h}
\end{align}
The remaining components correspond, in the appropriate fluid limits, to kinetic-helicity, cross-helicity, and divergence-related sectors.

\subsection{Kinetic Saffman integrals}\label{subsec:IIIC}
We now turn from the local conservation laws to the fluctuation integrals that can remain useful in globally nonhelical fields. Equations~\ref{eq:hLeqn1} and~\ref{eq:hL_l} provide strictly local conservation statements for $\mathscr{L}_{\iota}$ and $\mathscr{L}_{\iota, l}$, but in a globally nonhelical, sign-indefinite field, it does not follow that the corresponding domain-integrated quantities are useful. Indeed, for an initially random configuration the full-box integrals, 
\begin{subequations}
\begin{align}
\mathcal{L}_{V \iota} = \int_{V} d^d x \; \mathscr{L}_{\iota} (\mathbf{x}),
\end{align}
\begin{align}
\mathcal{L}_{V \iota, l} = \int_{V} d^d x \; \mathscr{L}_{\iota, l} (\mathbf{x}),
\end{align}
\end{subequations}
can vanish, or remain small, by cancellation even while individual structures contain finite $\mathscr{L}_{\iota}$ and $\mathscr{L}_{\iota, l}$. In such circumstances, the appropriate global measures are instead the typical fluctuation content of the locally conserved densities on scales large compared to an interaction region but small compared to the system size. Following the logic used for nonhelical MHD decay, we therefore consider a Saffman-type correlation integral, analogous to $\mathcal{I}_H$,
\begin{align}
\mathcal{I}_{\mathscr{L} \iota} (R) = \int_{V_R} d^d r \; \langle \mathscr{L}_{\iota} (\mathbf{x}) \mathscr{L}_{\iota} (\mathbf{x} + \mathbf{r}) \rangle, \label{eq:saffman1}
\end{align}
with $L^d \ll V_R \ll V_{\text{sys}}$, and related integrals for the conserved components of $\mathscr{L}_{\iota}$,
\begin{align}
\mathcal{I}_{\mathscr{L} \iota, l} (R) = \int_{V_R} d^d r \; \langle \mathscr{L}_{\iota, l} (\mathbf{x}) \mathscr{L}_{\iota, l} (\mathbf{x} + \mathbf{r}) \rangle. \label{eq:saffman2}
\end{align}
Should they exist, we denote the plateau values of $\mathcal{I}_{\mathscr{L} \iota} (R)$ and $\mathcal{I}_{\mathscr{L} \iota, l} (R)$ as $I_{\mathscr{L} \iota}$ and $I_{\mathscr{L} \iota, l}$, respectively. Under approximate homogeneity and isotropy, Eqs.~\ref{eq:hLeqn1} and~\ref{eq:hL_l} imply (see, e.g., Appendix B of~\citet{Hosking2021})
\begin{subequations}
\begin{align}
- \frac{1}{2} \frac{dI_{\mathscr{L} \iota}}{dt} = \int_{V_R} d^d r \; \nabla_{\mathbf{r}} \cdot \langle \mathscr{L}_{\iota} (\mathbf{x}) \mathbf{F}_{\iota} (\mathbf{x} + \mathbf{r}) \rangle, \label{subeq:ICresult}
\end{align}
\begin{align}
- \frac{1}{2} \frac{dI_{\mathscr{L} \iota, l}}{dt} = \int_{V_R} d^d r \; \nabla_{\mathbf{r}} \cdot \langle \mathscr{L}_{\iota, l} (\mathbf{x}) \mathbf{F}_{\iota, l} (\mathbf{x} + \mathbf{r}) \rangle, \label{subeq:IClresult}
\end{align}
\end{subequations}
whose $r.h.s.$ can vanish, i.e., there are intermediate-scale invariants that survive finite-box effects~\citep{Zhou2022}, if the $\mathbf{r}$-space flux terms have vanishing surface fluxes at large radii; under isotropy, this is ensured if the corresponding radial flux correlations each satisfy
\begin{subequations}
\label{eq:decorrelation}
\begin{align}
\hat{\mathbf{r}} \cdot 
\left\langle \mathscr{L}_{\iota}(\mathbf{x})
\mathbf{F}_{\iota}(\mathbf{x}+\mathbf{r}) \right\rangle
&= o(r^{1-d}),
\label{subeq:decorrelation_full}
\\
\hat{\mathbf{r}} \cdot 
\left\langle \mathscr{L}_{\iota,l}(\mathbf{x})
\mathbf{F}_{\iota,l}(\mathbf{x}+\mathbf{r}) \right\rangle
&= o(r^{1-d}),
\label{subeq:decorrelation_component}
\end{align}
\end{subequations}
for $L \ll r \ll L_{\text{sys}}$. 

We reason that the derivation here applies for 2D3V PIC, such as the numerical code we utilize for this work (see Appendix~\ref{appA}), because the source-compensated densities, $\mathscr{L}_{\iota}$ and $\mathscr{L}_{\iota, l}$, and the kinetic Saffman integral plateau values, $I_{\mathscr{L} \iota}$ and $I_{\mathscr{L} \iota, l}$, remain well defined in 2D systems with 3D fields and velocities that are planar (in $x$-$y$) isotropic, provided the same conditions used in 3D hold. Due to computational constraints, we exclusively consider 2D3V simulations and leave three-dimensional, three-velocity-component studies, with all three components of the electromagnetic fields retained, for future work. In our 2D3V runs we treat these assumptions as working hypotheses and assess them empirically via the emergence of $L \ll R \ll L_{\text{sys}}$ plateaus. 

Assuming ergodicity, i.e., that spatial sampling over our periodic domain is representative of an ensemble average~\citep[e.g.,][]{Hosking2021}, we may arrive at
\begin{subequations}
\begin{align}
I_{\mathscr{L} \iota} = \lim_{V \to \infty} \frac{1}{V} \langle \mathcal{L}_{V \iota}^2 \rangle, \label{subeq:ILiota_estimate}
\end{align}
\begin{align}
I_{\mathscr{L} \iota, l} = \lim_{V \to \infty} \frac{1}{V} \langle \mathcal{L}_{V \iota, l}^2 \rangle. \label{subeq:ILliota_estimate}
\end{align}
\end{subequations}
Through Eqs.~\ref{subeq:ILiota_estimate} and~\ref{subeq:ILliota_estimate}, $I_{\mathscr{L} \iota}$ and $I_{\mathscr{L} \iota, l}$ are gauge invariant under the same boundary/localization assumptions that render $\mathcal{L}_{V \iota}$ and $\mathcal{L}_{V \iota, l}$ meaningful.

\begin{figure*}
\begin{center}
\includegraphics[width=1.0\textwidth]{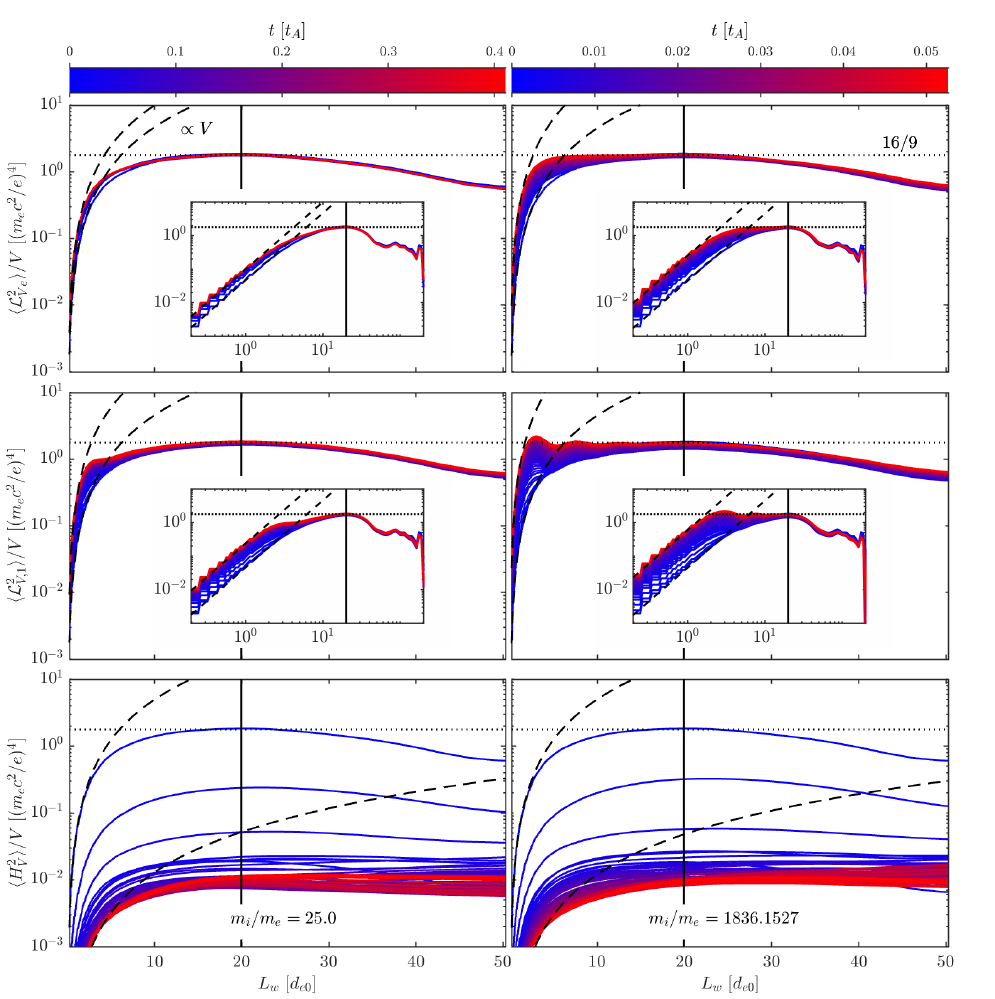}
\caption{\label{fig:demo} Demonstration of the approximate invariance of $I_{\mathscr{L} e}$ (upper row) and $I_{\mathscr{L}, 1}$ (middle row), calculated via $\langle \mathcal{L}_{V e}^2 \rangle/V$ and $\langle \mathcal{L}_{V, 1}^2 \rangle/V$, respectively, for $m_i/m_e = 25$ (left column) and $\approx 1836$ (right column). The lower row shows the estimator for $I_H$ ($\langle H_V^2 \rangle/V$), whose plateau values vary by over 2 orders of magnitude during the evolution. Included (dotted black) is the calculated value of $I_{\mathscr{L}, 1} (t = 0) = I_H (t = 0)$ ($=I_{\mathscr{L} e} (t = 0)$ for initially zero flow), derived in Appendix~\ref{appB}, and $\propto V = L_w^2$ curves (dashed black) for small-window sizes. The solid black lines show the window size at which we calculate $I_{\mathscr{L} e}$, $I_{\mathscr{L}, 1}$, and $I_H$, plotted in Fig.~\ref{fig:Saffman_evol_theory}. }
\end{center}
\end{figure*}

Focusing on the electron source-compensated integral, $I_{\mathscr{L} e}$, for subion dynamics, Fig.~\ref{fig:demo} shows the time evolution of $\langle \mathcal{L}_{V e}^2 \rangle/V$, $\langle \mathcal{L}_{V, 1}^2 \rangle/V$ (where $\mathcal{L}_{V,1}$ denotes the subvolume integral of the magnetic component $\mathscr{L}_1 \equiv \mathscr{L}_{\iota,1}$), and $\langle H_V^2 \rangle/V$~\footnote{We compute $\langle \mathcal{L}_{V e}^2 \rangle/V$, $\langle \mathcal{L}_{V, 1}^2 \rangle/V$, and $\langle H_V^2 \rangle/V$ from gridded 2D snapshots of $\mathscr{L}_e$, $\mathscr{L}_1$, and $h$ by sliding a rectangular window over all positions with periodic wraparound, summing the scalar fields within each window to form window-integral maps, or equivalently, circular convolutions with rectangular top-hat kernels. We then average the squared window integrals and normalize by the window area. For mean-zero stationary fields, this quantity is an area-weighted measure of the fields' two-point correlation within the chosen window scale.}, with $V = L_w^2$ the window area, from our globally nonhelical simulations. This figure demonstrates that $I_{\mathscr{L} e}$ and $I_{\mathscr{L}, 1}$, the estimated plateau values of $\langle \mathcal{L}_{V e}^2 \rangle/V$ and $\langle \mathcal{L}_{V, 1}^2 \rangle/V$, remain approximately constant over the interval and window sizes where $L \ll R \ll L_{\text{sys}}$~\footnote{Note that the decrease in $\langle \mathcal{L}_{V e}^2 \rangle/V$ and $\langle \mathcal{L}_{V, 1}^2 \rangle/V$ at large $L_w$ in each panel of Fig.~\ref{fig:demo} are finite-size artifacts, as the full-box integral is exactly zero by construction.} plateaus are identifiable, e.g., compared to the evolution of $I_H$ depicted in both Figs.~\ref{fig:demo} and~\ref{fig:Saffman_evol_theory}, the latter plotting the approximate plateau values in Fig.~\ref{fig:demo}. Although Fig.~\ref{fig:Saffman_evol_theory} indicates that both $I_{\mathscr{L} e}$ and $I_{\mathscr{L}, 1}$ demonstrate near constancy, $I_{\mathscr{L}, 1}$ appears slightly less invariant than $I_{\mathscr{L} e}$ and, as shown in Fig.~\ref{fig:demo}, $\langle \mathcal{L}_{V, 1}^2 \rangle/V$ seems to exhibit more variation at small scales than $\langle \mathcal{L}_{V e}^2 \rangle/V$ for both mass ratios simulated. $I_{\mathscr{L} e}$ likely appears better conserved because the additional kinetic components in $\mathscr{L}_e = \sum_l \mathscr{L}_{e,l}$ may introduce rapidly decorrelating cross terms that partially cancel the longer-range $\mathscr{L}_1$ contributions in $\hat{\mathbf{r}} \cdot \langle \mathscr{L}_e (\mathbf{x}) \mathbf{F}_e (\mathbf{x} + \mathbf{r}) \rangle$, consistent with two-fluid/extended MHD treatments in which the canonical helicity construction is often better suited for interpreting apparent magnetic-, cross-, and kinetic-helicity-like nonconservation as internally exchanging components~\citep{You2012,You2016,vonderLinden2018}. 

The fact that approximately time-independent plateaus show up in these plots is consistent with $\mathscr{L}_e$ and $\mathscr{L}_1$ having finite effective correlation lengths over the measured interval and that subvolume boundary fluxes that would change $\langle \mathcal{L}_{V e}^2 \rangle/V$ and $\langle \mathcal{L}_{V, 1}^2 \rangle/V$ are not building up system-spanning correlations; i.e., Eq.~\ref{eq:decorrelation} is satisfied. This is the same decorrelation assumption that underpins conservation of $I_H$ in nonhelical MHD~\citep[e.g.,][]{Hosking2021,Hew2026}, now applied to $I_{\mathscr{L} e}$ and $I_{\mathscr{L}, 1}$ in our case. 

Immediately evident in Fig.~\ref{fig:demo} is that although the plateau values of $\langle \mathcal{L}_{V e}^2 \rangle/V$ and $\langle \mathcal{L}_{V, 1}^2 \rangle/V$ remain approximately constant throughout both globally nonhelical simulations, the small-window component of the curves exhibits a slight upward shift during the early stages of the simulations, more so for the realistic mass-ratio run. We interpret this shift as an increase in the local variance/intensity of $\mathscr{L}_e$ and $\mathscr{L}_1$ at small $R$. Because there exists a much wider separation between ion and electron scales with $m_i/m_e \approx 1836$ compared to the $m_i/m_e = 25$ case, the cascade possibly has room to develop a more pronounced electron-scale range that can raise $\langle \mathcal{L}_{V e}^2 \rangle/V$ and $\langle \mathcal{L}_{V, 1}^2 \rangle/V$ at small $R$ over time, even while the large-window fluctuation content stays fixed, consistent with the behavior of a rugged constraint that governs large-scale/self-similar evolution, while small-scale physics controls intermittency and dissipation-range details~\citep{Kolmogorov1991,Goldreich1995,Frisch1995}. 

It is worth noting that, as a direct consequence of the reduction from Eq.~\ref{eq:reduction_h}, the first kinetic Saffman integral (or the ``magnetic integral''), $\mathcal{I}_{\mathscr{L}, 1} (R)$, in the ideal-MHD limit reduces to the Saffman magnetic-helicity integral,
\begin{align}
\mathcal{I}_{\mathscr{L}, 1} (R) \underset{\text{ideal MHD}}{\to} \mathcal{I}_H (R).
\end{align}
To demonstrate this point, we show in Fig.~\ref{fig:Saffman_evol_theory} the time evolution of $I_H$, which indicates that during late stages of the evolution, when the peak of the magnetic-energy spectrum reaches scales beyond $d_e$, occurring at times $t \gtrsim 0.22 \; t_A \approx 8.8 \; \Omega_{ci,0}^{-1}$ for $m_i/m_e = 25$ and $t \gtrsim 0.033 \; t_A \approx 0.15 \; \Omega_{ci,0}^{-1}$ for $m_i/m_e \approx 1836$, $I_H$ becomes better conserved as $\mathbf{E} \cdot \mathbf{B}$ becomes comparatively small and the time-history contribution evolves much more slowly and/or when $\mathbf{E} \cdot \mathbf{B}$ becomes statistically uncorrelated with structure-handedness proxies, as explored in Sec.~\ref{sec:Hevol}. This transition results in $I_H$ approaching a constant offset set by the earlier kinetic stage.

Here, the role of $\mathscr{L}_1$ is interval based rather than tied to a single event. Over the measured kinetic interval, its usefulness is assessed empirically by whether the corresponding kinetic Saffman integral develops an approximately time-independent intermediate-scale plateau. 

From the observed invariance of the kinetic Saffman integrals, we can extract a scaling constraint by connecting the magnetic integral construction to the macroscopic similarity variables measured in our simulations. As discussed in Sec.~\ref{sec:Hevol}, early-time interactions occur in a regime where nonideality depletes magnetic helicity while remaining closely tied to magnetic energy conversion (see, e.g., Eq.~\ref{eq:Hscalingeqn}). This motivated introducing $\mathscr{L}_1$ as an event-local helicity-balance density (Eqs.~\ref{eq:eventlocal} and~\ref{subeq:L_1}) for short interactions, and as the magnetic component of the source-compensated construction whose cumulative correlation integral is tested over the longer kinetic interval. We then made the working assumption—motivated by single-scale structure geometry and by the empirical association of the interaction length $L$ with both magnetic-helicity depletion and energy conversion—that the characteristic amplitude of $\mathscr{L}_1$ in an interaction scales as $\mathscr{L}_1 \sim B^2 L$ (Eq.~\ref{eq:workinghyp})
up to order-unity, event-dependent factors. Substituting this estimate into the Saffman integral scaling, e.g., Eq.~\ref{eq:saffman2}, and assuming that $\mathscr{L}_1$ structures are sufficiently localized, e.g., $\int d^d r \; | \langle \mathscr{L}_1 (\mathbf{x}) \mathscr{L}_1 (\mathbf{x} + \mathbf{r}) \rangle | < \infty$ so that the effective correlation volume is $V_{\text{corr}} \sim L^d$, using $d = 2$ for the present 2D3V geometry gives
\begin{align}
I_{\mathscr{L}, 1}^{1/4} \sim [(B^2 L)^2 V_{\text{corr}}]^{1/4} \sim B L \sim \text{const}, \label{eq:L1scaling}
\end{align}
from approximate invariance of $I_{\mathscr{L}, 1}$~\footnote{The scaling in Eq.~\ref{eq:L1scaling} is also consistent with random-walk arguments for structure coalescence, or equivalently with the random-walk interpretation of the subvolume-variance form of Eq.~\ref{subeq:ILliota_estimate}, in the absence of a nonzero mean helicity~\citep[e.g.,][]{Hosking2021}.}. Equation~\ref{eq:L1scaling} therefore provides a plausible mechanism—under approximate single-scale self-similarity—by which the $BL \sim \text{const}$ constraint can persist in our decaying 2D3V PIC regime, even though the usual 2D MHD invariant arguments that lead to $BL \sim \text{const}$ do not directly apply. For example, in 2D MHD turbulence, the scaling $BL \sim \text{const}$ follows directly from writing Eq.~\ref{eq:Azdef}. In the ideal limit, the mean-square out-of-plane magnetic vector potential $\int d^2 x \; |A_z|^2$, sometimes termed anastrophy, is a rugged invariant~\citep{Fyfe1976,Fyfe1977,Matthaeus1980,Hatori1984,Ting1986,Biskamp1989,Biskamp2001}, and since dimensionally $A_{z,\text{rms}} \sim BL$, approximate conservation of $\int d^2 x \; |A_z|^2$ implies $BL \sim \text{const}$~\citep{Hosking2021}. In fully kinetic 2D3V systems, however, $\int d^2 x \; |A_z|^2$ is generally not conserved because phase-space cascades, Landau damping, and finite-Larmor-radius effects introduce additional channels~\footnote{In 2D3V, the out-of-plane magnetic vector potential $A_z$ evolves as $\partial_t A_z = - c E_z$ and hence the evolution equation for anastrophy $\mathcal{A} \equiv \int d^2 x \; |A_z|^2$ includes a source term $-2 c \int d^2 x \; A_z E_z$. We find in the numerical experiments presented in this work that $\mathcal{A}$ is generally not conserved, even during late stages of the system evolution (not shown in this paper).}. Here, we have presented a fully kinetic argument for the constancy of $BL$ in 2D3V without appealing to frozen-in flux or to 2D MHD anastrophy conservation. 

\begin{figure}[b]
\includegraphics[width=0.5\textwidth]{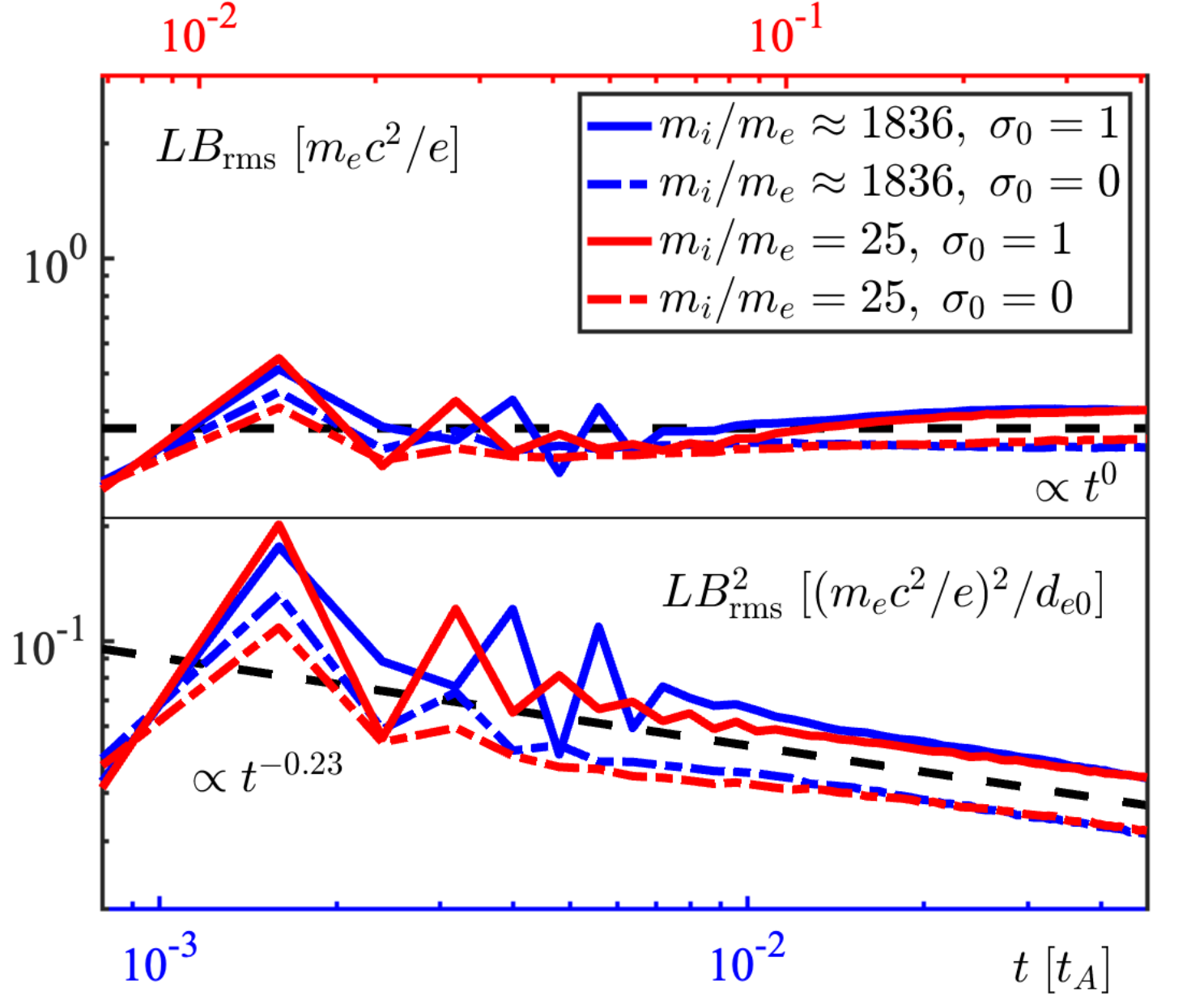}
\caption{\label{fig:BL} Numerical verification of $BL \sim \text{const}$ (Eq.~\ref{eq:L1scaling}) over all 2D3V PIC simulations (upper panel). The evolution of $B^2L$ (lower panel) is included for reference. The black dashed lines plot the power-law scalings: $\propto t^0$ in the upper panel and $\propto t^{-0.23}$ in the lower panel, determined from a linear least-squares fit of the $m_i/m_e \approx 1836$ and $\sigma_0 = 0$ run. }
\end{figure}

Applying the integral scale definition~\citep[e.g.,][]{Zhou2022}
\begin{align}
L \equiv \frac{\int_0^{\infty} dk \; k^{-1} \mathcal{E}_B (k)}{\int_0^{\infty} dk \; \mathcal{E}_B (k)}, \label{eq:intscale}
\end{align}
we show numerical evidence in agreement with the $BL \sim \text{const}$ scaling in Fig.~\ref{fig:BL}, which plots both $BL$ and $B^2L$, with the latter monotonically decreasing on average, for both initially net-helical and nonhelical simulations. Although the arguments presented in this section may provide an explanation as to why we find that $BL \sim \text{const}$ for our initially nonhelical runs, it is interesting that Fig.~\ref{fig:BL} should also show this to be the case in our initially net-helical simulations. We explore this point of contention in the following section. 

\section{Initially net-helical field}\label{sec:nethelical}
Let us now consider an initially net-helical configuration as a useful contrasting case for the scaling arguments leading to the $BL \sim \text{const}$ constraint. Since $\mathscr{L}_1$ provides a conserved diagnostic and scales as $\sim B^2 L$ up to order-unity, event-dependent factors, under a straightforward extension of the MHD intuition, one might be tempted to expect that with a net-helical initial configuration, the decay should obey the scaling $B^2 L \sim \text{const}$, e.g., the same scaling as in MHD for net-helical fields~\citep{Field2000,Christensson2001,Caprini2014}. In our numerical simulations with globally helical initial conditions, however, we find that although the configurations start with $\sigma_0 = 1$, the fields rapidly tend toward effectively nonhelical global states, i.e., $\sigma \to 0$. Thus, in the collisionless, initially net-helical 2D3V cases studied here, the spectral/windowed helicity diagnostics do not show MHD-like conservation over the early subion kinetic interval; instead, the initially single-signed helical content rapidly becomes cancellation dominated.

\begin{figure*}
\begin{center}
\includegraphics[width=1.0\textwidth]{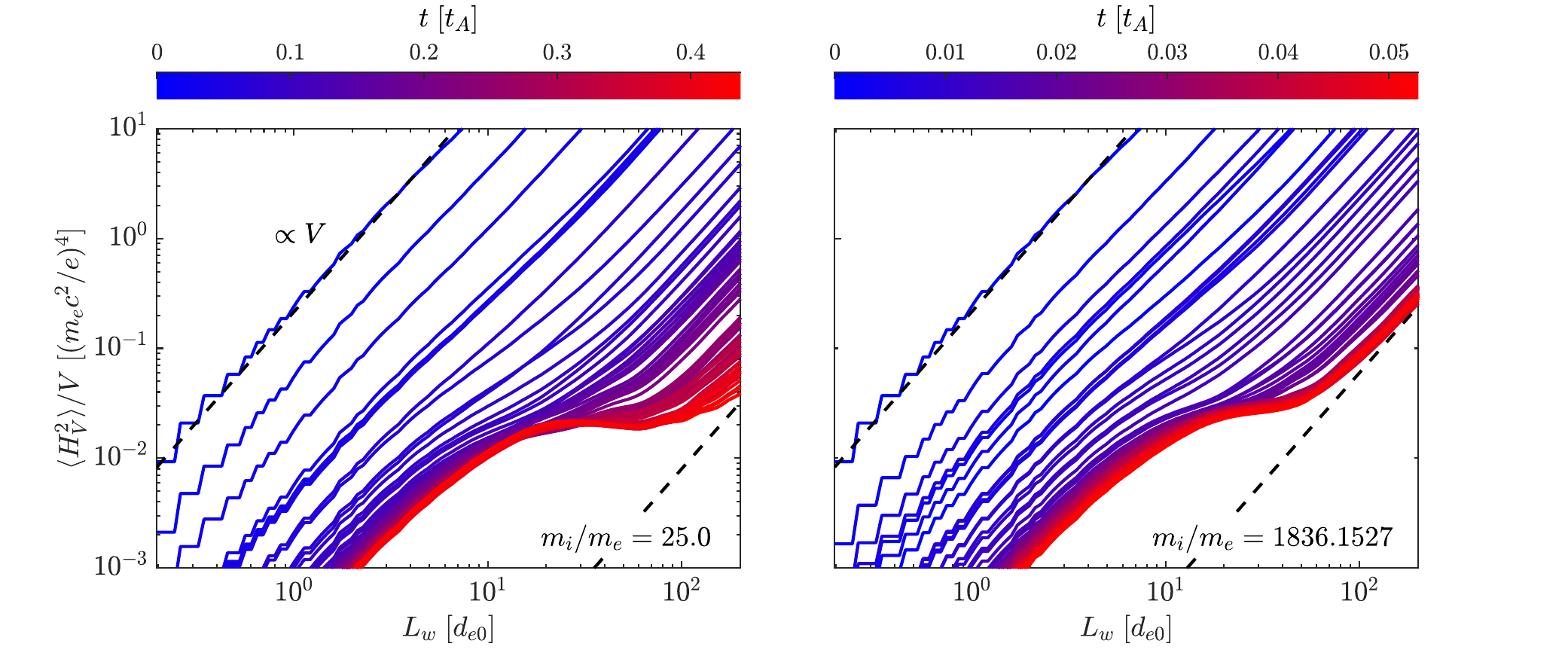}
\caption{\label{fig:helical_saff} Demonstration of the tendency toward zero global fractional helicity for initially net-helical configurations in both $m_i/m_e = 25$ (left panel) and $\approx 1836$ (right panel) runs. Included (dashed black) are $\propto V = L_w^2$ lines at small and large window sizes, showing the behavior of $\langle H_V^2 \rangle/V$ within individual magnetic structures and the small global helicity imbalance over larger windows. }
\end{center}
\end{figure*}

This trend is illustrated in Fig.~\ref{fig:helical_saff}, which shows, for our initially fully helical fields, the evolution of the windowed magnetic-helicity variance $\langle H_V^2\rangle/V$ that, at early times, exhibits an approximately $\propto V$ scaling across window sizes, consistent with magnetic helicity being predominantly single-signed over the accessible scales~\citep{Hosking2021}. As the evolution proceeds, $\langle H_V^2\rangle/V$ develops a plateau over a growing range of window sizes, indicative of increasing cancellation among positive/negative contributions within those windows, and a weak residual imbalance appears to remain primarily at the largest windows.

The emergence of mixed-handed substructures is consistent with the subion nonideality mechanism discussed in Sec.~\ref{sec:Hevol}. In our simulations, $\mathbf{E}\cdot\mathbf{B}\neq 0$ is intermittent and localized, so different regions experience different histories of the integrated nonideal term $- 2c \int_{t_0}^t dt^{\prime} \; \mathbf{E} \cdot \mathbf{B}$. In regions where this term becomes comparable to or exceeds the local preexisting helical content, the local magnetic-helicity density can be driven through zero and acquire the opposite sign. Consistent with Sec.~\ref{sec:Hevol}, we find that $\text{sgn}(\mathbf{E}\cdot\mathbf{B})$ predominantly agrees with $\text{sgn}(\mathbf{E}^{\prime} \cdot\mathbf{J}/\alpha)$ across the cases examined, as demonstrated in Fig.~\ref{fig:EJsign}, so the same correlation between nonideality and structure handedness is observed within the same empirical scatter in the initially net-helical runs as well.

Since the global fractional helicity rapidly becomes increasingly cancellation dominated, the data are most consistent with the nonhelical scaling $BL \sim \text{const}$ over the subion interval we measure, as reported by Fig.~\ref{fig:BL}. At early times, there is a brief transient in which the $BL$ curve is slightly elevated relative to the $\sigma_0 = 0$ case, consistent with an intermediate regime while the field transitions from single-signed to cancellation-dominated magnetic helicity. This transient may then contribute to slightly larger final $LB_{\text{rms}}$ and $LB_{\text{rms}}^2$ compared to the initially nonhelical runs, as indicated in Fig.~\ref{fig:BL}.

\section{Conclusions}\label{sec:conclusions}
Magnetic helicity has long served as a powerful organizing principle for magnetically dominated turbulence in ideal MHD, where it is approximately conserved up to boundary fluxes and weak resistive effects and therefore constrains large-scale evolution and inverse transfer. Many space and astrophysical plasmas, however, transition into regimes where ion and electron dynamics decouple, current sheets thin to kinetic scales, and collisionless reconnection becomes an essential part of the cascade and dissipation. A central result of this work is that, in the class of subion decaying turbulence studied here, helicity-based reasoning can be retained if reformulated in a fully kinetic manner that explicitly accounts for localized nonideality.

Taken together, these results suggest that the significance of the present study is not limited to the particular 2D3V PIC realization considered here, but lies in identifying a kinetic mechanism and a corresponding diagnostic framework by which helicity-constrained decay can remain informative even when ideal-MHD magnetic-helicity conservation is locally compromised.

Directly from magnetic-helicity balance, the volume term proportional to $\int_V dV \; \mathbf{E} \cdot \mathbf{B}$ provides a channel for magnetic-helicity change whenever parallel electric fields appear, as is commonly observed and expected in collisionless diffusion regions. In our 2D3V PIC simulations, $\mathbf{E} \cdot \mathbf{B} \neq 0$ is intermittent and spatially localized, and, during the early kinetic stage, shows a statistically significant association with structure handedness, diagnosed via fixed-gauge magnetic- and current-helicity proxies. This behavior is consistent with reconnection-mediated interactions that tend to decrease $|H_{V_s}|$, the structure-integrated magnetic-helicity diagnostic in the specified gauge, and that coincide with a decline of the usual Saffman helicity-variance plateau value $I_H$.

Motivated by the observed role of time-integrated $\mathbf{E} \cdot \mathbf{B}$, we revisited canonical vorticity transport starting from the Vlasov-Maxwell system and derived an exact local continuity equation with a kinetic source term. By absorbing the time-integrated source into a redefined, history-dependent density, we obtained a source-free conservation identity by construction and Saffman-type correlation integrals, which, under standard flux-decorrelation assumptions, can admit intermediate-scale plateaus that are approximately time independent. The source-compensated quantity is an exact reformulation by construction; the physically relevant question is whether intermediate-scale plateaus emerge empirically under localization assumptions. Once an approximately conserved plateau exists, it can play an analogous role as the Saffman magnetic-helicity integral plays in MHD decay theory. In the globally nonhelical case, the first kinetic Saffman integral constrains the coupled evolution of magnetic field amplitude and outer scale under approximate single-scale self-similarity. In our 2D3V setting, invariance of the kinetic integral is consistent with the familiar nonhelical scaling $BL \sim \text{const}$, and our simulations show behavior compatible with this scaling over the measured subion interval. Within this framework, reconnection enters explicitly through the time-history term, so one can formulate decay constraints that remain informative across the kinetic interval when a plateau is observed. The interpretation of the source-compensated quantity depends on the chosen reference interval. When the reference time is tied to a particular short subion interaction, it provides an event-local balance diagnostic; when applied over the broader kinetic interval, it should instead be understood as a cumulative quantity whose usefulness is assessed through the observed plateau behavior of its correlation integral.

In our simulations, initially net-helical configurations quickly develop mixed-signed magnetic-helicity patches and the global fractional helicity decreases toward small values. This behavior is consistent with, and plausibly driven by, the intermittency of $\mathbf{E} \cdot \mathbf{B}$, whereby different regions accumulate different histories of the nonideal term, so the local magnetic-helicity density can be driven through zero and change sign. The emergence of strong sign mixing is qualitatively consistent with the modern observational and numerical picture of kinetic turbulence as an intermittent ensemble of reconnecting structures, in which reconnection and dissipation signatures cluster around current sheets. One implication is that the presence of a net-helical bias in the initial spectrum does not automatically enforce the MHD-like $B^2L \sim \text{const}$ scaling over the kinetic interval examined in our 2D3V runs. Instead, kinetic reconnection can rapidly reduce the net magnetic helicity before an MHD-scale helicity constraint becomes relevant.

In weakly collisional plasmas, one possible implication is that subion reconnection may act as a sign-mixing mechanism that reduces the net magnetic helicity available to be inherited by larger, more MHD-like scales unless magnetic helicity is continuously replenished by large-scale driving or boundaries. This possibility is at least qualitatively relevant to environments such as the solar wind and magnetosheath, where turbulence spans MHD and kinetic scales and exhibits abundant reconnection signatures and intermittent dissipation, and it may also motivate caution when applying helicity-constrained decay arguments in more speculative astrophysical settings where a portion of the magnetic spectrum enters non-MHD regimes~\citep{Grasso2001,Banerjee2004,Durrer2013,Caprini2014,Subramanian2016}. A related possibility is that collisionless nonideality provides a kinetic-scale pathway for reducing the helical content of small-scale magnetic structures, which may be relevant to longstanding questions about small-scale helicity accumulation and dynamo saturation~\citep{Pouquet1976,Gruzinov1994,Cattaneo1996,Vishniac2001,Field2002,Blackman2002,Brandenburg2005_b}. These connections are necessarily speculative in our 2D3V, freely decaying setting, but they point to concrete tests in 3D kinetic and hybrid-kinetic systems.

The most valuable next steps are those that probe robustness, universality, and measurability. The existence of an intermediate-scale plateau—and the localization assumptions needed for conservation—should be tested in fully 3D kinetic turbulence. Extending the framework from freely decaying to forced turbulence would clarify whether the conserved integral constrains statistically steady-state cascades, e.g., by setting relationships between injection, intermittency, and outer scale evolution, in the presence of sustained reconnection. Systematic scans in $\beta$, guide-field strength, imbalance, and scale separation can determine when the tendency toward mixed-signed magnetic-helicity patches is fast across a broad range of parameters, versus when net magnetic helicity can survive the kinetic range.

\section*{Acknowledgments}\label{sec:acknowledgments}
The author extends deep gratitude to the late Nuno F. Loureiro for insightful discussions, for immense support during the early stages of this work, and for suggesting the appropriate symbol to represent the locally conserved kinetic density $\mathscr{L}_{\iota}$. The author thanks the anonymous referees for suggestions that significantly improved this paper. The author also acknowledges the OSIRIS Consortium, comprising UCLA and Instituto Superior Técnico (IST) in Lisbon, Portugal, for providing access to the OSIRIS 4.0 framework.

\section*{Funding}\label{sec:funding}
This research was partially funded by DOE award DE-FG02-91ER54109 and utilized resources of the National Energy Research Scientific Computing Center, a DOE Office of Science User Facility, supported by the U.S. Department of Energy under Contract No. DE-AC02-05CH11231 using NERSC award FES-ERCAP0026577. This material is based upon work supported by the National Science Foundation Graduate Research Fellowship Program under Grant No. 2141064. Any opinions, findings, and conclusions or recommendations expressed in this material are those of the author and do not necessarily reflect the views of the National Science Foundation.

\section*{Data Availability}
There are no publicly available research data or software supporting this manuscript. Requests for further information or data should be sent to the author.

\appendix

\section{Simulation setup and diagnostic details}\label{appA}
The simulations reported in this work are conducted using the PIC code \textsc{osiris}~\citep{Fonseca2002,Hemker2015}. The global Alfv\'en time is defined as $t_A = L_x/v_{A,i0}$~\footnote{For reference, the global Alfv\'en time may be related to the initial ion cyclotron time through $t_A \Omega_{ci,0} = L_x/d_{i0} = (L_x/d_{e0}) \sqrt{m_e/m_i}$, where $\Omega_{ci,0} = e B_{\text{rms},0}/(m_i c)$ and $d_{i0} = c/\omega_{pi}$. Because $L_x/d_{e0} = 64\pi$ is fixed in all runs, this gives $t_A \Omega_{ci,0} \approx 40.2$ for $m_i/m_e = 25$ and $t_A \Omega_{ci,0} \approx 4.69$ for $m_i/m_e \approx 1836$.}, where $v_{A,i0} = B_{\text{rms}, 0}/(4\pi m_i n_0)^{1/2}$ is the ion Alfv\'en speed, $B_{\text{rms}, 0} \equiv B_{\text{rms}} (t = 0)$, and $L_x$ and $L_y$ are the lengths of the simulation box in the $x$ and $y$ directions, respectively, and are equivalent. The initial ion and electron temperatures $T_{i0}$ and $T_{e0}$ are equal and uniform, we set $\beta_{\text{rms}} \equiv 8\pi n_0 (T_{i0} + T_{e0})/B_{\text{rms}, 0}^2 = 1$, and species' bulk flows are initialized as zero. The relationship between the electron plasma and cyclotron frequencies is given by $\omega_{pe} = 2\Omega_{ce,\text{rms}}$ where $\omega_{pe} = (4\pi n_0 e^2/m_e)^{1/2}$ and $\Omega_{ce,\text{rms}} = |e|B_{\text{rms}, 0}/(m_e c)$. We utilize quadratic particle interpolations and resolve the electron Debye length, $\lambda_{De} = u_{\text{th},e}/\omega_{pe}$, to mitigate numerical heating.

To compare cancellation-dominated and net-helical initial conditions, we parametrize the initial magnetic field using the global fractional magnetic helicity
\begin{align}
\sigma = \frac{\int_0^{\infty} dk \; \sigma_k (k)\mathcal{E}_B (k)}{\int_0^{\infty} dk \; \mathcal{E}_B (k)} \in [-1,1],
\label{eq:frachelicity}
\end{align}
with $\sigma_k (k) \equiv k H_k (k)/\mathcal{E}_B (k) \in [-1,1]$ the fractional magnetic-helicity spectrum~\citep[e.g.,][]{Matthaeus1982_a,Matthaeus1982_b,Howes2010}, and $\mathcal{E}_B (k)$ and $H_k (k)$ the magnetic energy and helicity spectra, respectively. In the simulations presented here, we consider $\sigma_0 \equiv \sigma(t = 0) = \{ 0,1 \}$. An alternative global normalization would be
\begin{align}
\sigma_L \equiv \frac{H}{L E_B}
=
\frac{\int_0^{\infty} dk \; k^{-1}\sigma_k(k)\mathcal{E}_B(k)}
{\int_0^{\infty} dk \; k^{-1}\mathcal{E}_B(k)} ,
\end{align}
where $H = \int dk \; H_k(k)$, $E_B = \int dk \; \mathcal{E}_B(k)$, and $L$ is the integral scale defined in Eq.~\ref{eq:intscale}. Thus, $\sigma_L$ is a $k^{-1}\mathcal{E}_B$-weighted average of $\sigma_k$, whereas Eq.~\ref{eq:frachelicity} is an $\mathcal{E}_B$-weighted average. The two definitions coincide when $\sigma_k$ is scale independent, and are approximately equivalent for narrow-band or single-scale spectra. We retain Eq.~\ref{eq:frachelicity} because it directly parametrizes the spectral helicity polarization used in the initial conditions; for the present initial conditions, $\sigma_k(k)=\sigma_0$, so both definitions give $\sigma_L(t=0)=\sigma(t=0)=\sigma_0$. In more broadband states, the difference between the two definitions can be useful: $\sigma_L$ emphasizes the larger-scale contribution to the net helicity, while Eq.~\ref{eq:frachelicity} emphasizes the energy-weighted mean polarization of the magnetic spectrum.

Because fully kinetic turbulence spans disparate ion and electron scales, we perform 2D3V PIC runs at both reduced ($m_i/m_e = 25$) and realistic ($m_i/m_e \approx 1836$) mass ratios. The reduced-ratio runs provide greater dynamical range and statistical access at finite computational cost, while the realistic-ratio runs verify that the main behaviors persist when the physical ion-electron-scale separation is retained. 

Key simulation parameters used in these runs are listed in Table~\ref{tab:table1}.

\begin{table}
\begin{ruledtabular}
\begin{tabular}{cccc}
& Parameter & Value & \\
\hline
& $\lambda_{De}/d_{e0} = u_{\text{th}, e}/c$ & $1/4$ & \\
& $L_x/d_{e0} = L_y/d_{e0}$ & $64 \pi$ & \\
& $\omega_{pe}/\Omega_{ce,\text{rms}}$ & $2$ & \\
& $T_{i0}/T_{e0}$ & $1$ & \\
& $m_i/m_e$ & $25.0$, $1836.1527$ \\
& $k_c d_{e0}$ & $0.5$ & \\
& $\beta_{\text{rms}}$ & $1$ & \\
& $M$ & $15$ & \\
& $\sigma_0$ & $0, 1$ & \\
& $a$ & $7$ & \\
\hline
& $\Delta x/\lambda_{De} = \Delta y/\lambda_{De}$ & $0.393$ & \\
& Time step ($\omega_{pe} \Delta t$) & $0.0687$ & \\
& Particles per cell (per species) & $4096~(64^2)$ & \\
& Resolution & $2048^2$ & \\
\end{tabular}
\caption{Summary of 2D3V simulation parameters.}
\label{tab:table1}
\end{ruledtabular}
\end{table}

Our magnetic field initialization is similar to that of~\citet{Hosking2021}, but the practical implementation differs. Because \textsc{osiris} specifies initial conditions via real-space expressions, we initialize $\mathbf{B}$ as a finite trigonometric sum, limiting the number of modes without compromising the robustness of parsing and evaluation.

Mode amplitudes are weighted by a prescribed shape function
\begin{align}
F_0(k) \equiv F (k; t = 0) = A k^{(a - d + 1)/2} \nonumber \\
\times \exp (1 - k^2/k_c^2), \label{eq:originalFk}
\end{align}
where $a$ is a tunable parameter used to place the excited band in $k$ space and $k_c$ sets its characteristic scale. Because only a narrow band of $k$ is populated with a finite number of modes and the field is subsequently renormalized to a prescribed $B_{\text{rms}, 0}$, $a$ should be viewed as a parameter controlling the initial excitation scale rather than as an independently realized inertial-range exponent. 

To obtain an approximately isotropic realization, we write discrete wave vectors $\mathbf{k}_m = ( 2\pi m_x/L_x, 2\pi m_y/L_y )$ and choose $M = 15$ distinct lattice pairs $(m_x, m_y)$, where $m_x, m_y \in \mathbb{Z}$, excluding the zero mode $(0, 0)$. For sparse-mode initialization, we center the populated ring near the wave number $|\mathbf{k}| \simeq k_p$ that maximizes the quadratic magnetic weight $k^{d - 1} F_0^2 (k)$, which controls random-phase estimates of large-scale fluctuations of quadratic magnetic quantities, e.g., $B^2$ and Maxwell stress. For Eq.~\ref{eq:originalFk}, this occurs at $k_p = \sqrt{a/4} k_c$~\footnote{The helicity-variance integral that defines $I_H$ carries an additional factor $k^{-2}$ (e.g., Eq.~\ref{eq:I_H_variancedef}), shifting the dominant band to $|\mathbf{k}| = \sqrt{(a - 2)/4} k_c$; this differs from $k_p$ only by $\approx 15 \%$ for $a = 7$ used here. For $d = 2$, $k_p$ differs from $\sqrt{(a + d - 1)/4} k_c$, the peak of the initial shell-integrated magnetic-energy spectrum $\propto k^{d - 1} F_0 (k)$, by $\approx 7 \%$ and from the reciprocal of the initial effective magnetic-helicity-density correlation length, $L_h^{-1} (t = 0)$, (i.e., Eq.~\ref{eq:emhdcl}) by $\approx 4 \%$.}.

We set the weights
\begin{align}
w_{\pm} [\sigma_k (k)] = [1 \pm \sigma_k (k)]/2,
\end{align}
draw independent phases $\phi_{\pm} \in \text{Unif} [0, 2\pi)$, and set the Fourier representation of the magnetic field as
\begin{align}
\mathbf{B}_{\mathbf{k}} = \sqrt{F_0(k)} [\sqrt{w_+ (\sigma_0)} e^{i\phi_+} \hat{\mathbf{e}}_+ \nonumber \\
+ \sqrt{w_- (\sigma_0)} e^{i\phi_-} \hat{\mathbf{e}}_-], \label{eq:Frep}
\end{align}
with the helical (circular) basis
\begin{align}
\hat{\mathbf{e}}_{\pm} = (\hat{\mathbf{e}}_1 \pm i \hat{\mathbf{e}}_2)/\sqrt{2},
\end{align}
such that $\hat{\mathbf{e}}_1 \perp \hat{\mathbf{k}}_m$ and $\hat{\mathbf{e}}_2 = \hat{\mathbf{k}}_m \times \hat{\mathbf{e}}_1$, i.e., each mode is divergence-free and $(\hat{\mathbf{e}}_1, \hat{\mathbf{e}}_2, \hat{\mathbf{k}}_m)$ is right-handed. For each mode, the real-space contribution of $B_{j, \mathbf{k}_m}$ is
\begin{align}
B_{j,m} (\mathbf{x}) = \Re \{ B_{j, \mathbf{k}_m} e^{i \mathbf{k}_m \cdot \mathbf{x}} \} & \nonumber \\
= \alpha_{j,m} \cos (\mathbf{k}_m \cdot \mathbf{x}) & - \beta_{j,m} \sin (\mathbf{k}_m \cdot \mathbf{x}),
\end{align}
where $\alpha_{j,m} = \Re \{ B_{j, \mathbf{k}_m} \}$ and $\beta_{j,m} = \Im \{ B_{j, \mathbf{k}_m} \}$. Finally, we write
\begin{align}
B_j (x, y) = \sum_{m = 1}^M A_{j,m} \cos (k_{xm} x + k_{ym} y + \phi_{j,m}), \label{eq:cos_sum}
\end{align}
with
\begin{subequations}
\begin{align}
A_{j,m} = \sqrt{\alpha_{j,m}^2 + \beta_{j,m}^2},
\end{align}
\begin{align}
\phi_{j,m} = \text{atan}2 (\beta_{j,m}, \alpha_{j,m}).
\end{align}
\end{subequations}
Over a periodic box, the spatial average of $|\mathbf{B}|^2$ is
\begin{align}
\langle |\mathbf{B}|^2 \rangle_{V_{\text{sys}}} = \frac{1}{V_{\text{sys}}} \int d^2 x \; |\mathbf{B}|^2 = \sum_{m = 1}^M \sum_{j = 1}^3 \frac{A_{j,m}^2}{2}.
\end{align}
Thus, all coefficients are rescaled by $\sqrt{2} B_{\text{rms}, 0}/(\sum_{m = 1}^M \sum_{j = 1}^3 A_{j,m}^2)^{1/2}$.

In addition to the initialization procedure, several diagnostics used in the main text require postprocessing of the simulation outputs. For the diagnostics used in Sec.~\ref{sec:Hevol}, we reconstruct the magnetic vector potential from the magnetic field at each output time. On the doubly periodic domain, we obtain $\mathbf{A}(x,y) = (A_x, A_y, A_z)$ by spectral inversion of the Poisson problem
\begin{align}
\nabla^2 \mathbf{A} = - \nabla \times \mathbf{B},
\end{align}
or equivalently,
\begin{align}
\tilde{\mathbf{A}}(\mathbf{k}) = i\, \mathbf{k} \times \tilde{\mathbf{B}}(\mathbf{k}) / |\mathbf{k}|^2,
\end{align}
with $\mathbf{k} \neq \mathbf{0}$ and the mean mode fixed by $\tilde{\mathbf{A}} (\mathbf{0}) = \mathbf{0}$. This yields the periodic Coulomb-gauge representative, for which $\nabla \cdot \mathbf{A} = 0$, and removes the residual constant-offset freedom.

\begin{figure*}
\begin{center}
\includegraphics[width=1.0\textwidth]{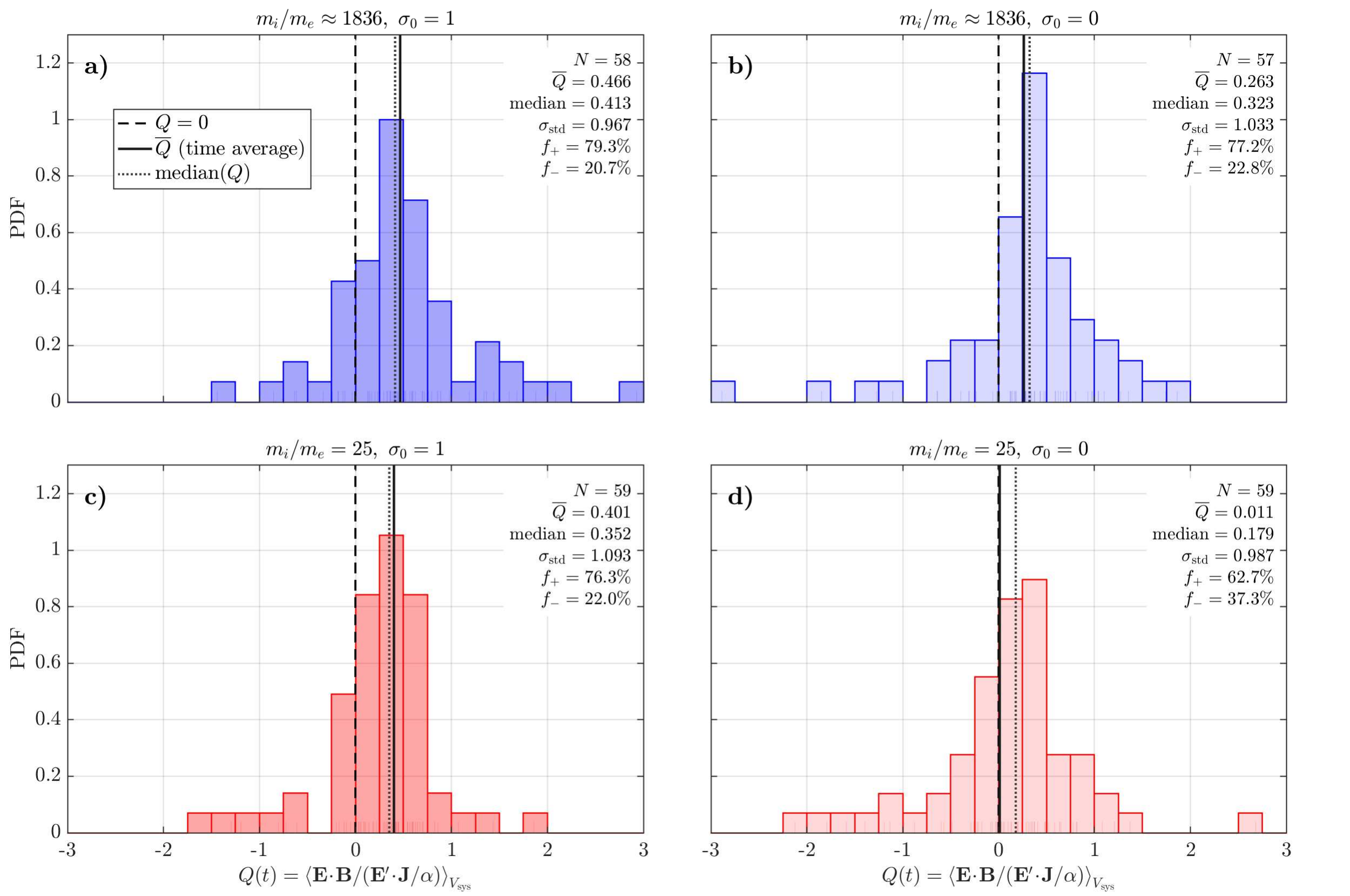}
\caption{\label{fig:EJsign_hist_app} Histograms over sampled output times of the domain-averaged ratio $Q(t) \equiv \left\langle \mathbf{E}\cdot\mathbf{B}/(\mathbf{E}^{\prime}\cdot\mathbf{J}/\alpha)\right\rangle_{V_{\text{sys}}}$ for the four 2D3V PIC simulations considered in this work. The dashed vertical line marks $Q=0$, the solid vertical line marks the time-average value $\overline{Q}$, and the dotted vertical line marks the median of the sampled distribution. Each panel also reports the number of sampled outputs, the mean, the median, the standard deviation, and the fractions $f_{+}$ and $f_{-}$ of sampled times for which $Q(t)$ is positive and negative, respectively. The distributions are predominantly positive in all four runs, while intermittent negative domain averages remain present.}
\end{center}
\end{figure*}

As an additional diagnostic relevant to the empirical relation in Sec.~\ref{sec:Hevol}, we also examine the distribution over sampled output times of the domain-averaged ratio $Q(t)$ (Eq.~\ref{eq:ratio}). Figure~\ref{fig:EJsign_hist_app} shows histograms of $Q(t)$ for each simulation run, normalized to represent probability density functions (PDFs), so that the total area under each histogram is unity. These distributions provide a compact summary of the time traces shown in Fig.~\ref{fig:EJsign}(a): In all four runs, the sampled values remain concentrated within an order-unity range and are predominantly positive, although negative domain averages occur intermittently. The vertical reference lines indicate $Q = 0$, the time-average value $\overline{Q}$, and the median value of the sampled distribution in each run.

\begin{figure}
\includegraphics[width=0.5\textwidth]{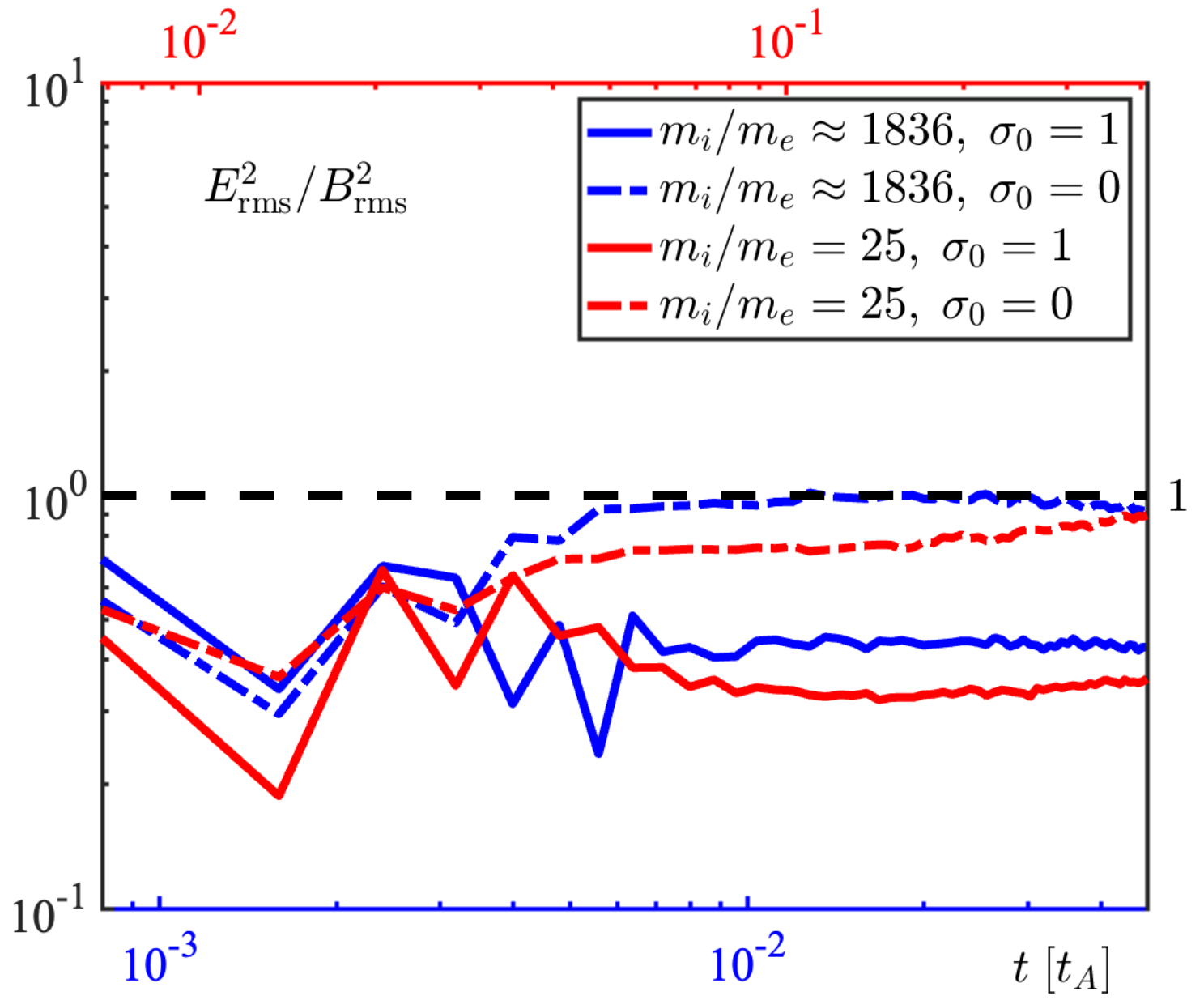}
\caption{\label{fig:EvsB_app} Evolution of $E_{\text{rms}}^2/B_{\text{rms}}^2$ for all 2D3V simulations presented in this work. The black dashed line indicates $E_{\text{rms}}^2/B_{\text{rms}}^2 = 1$.}
\end{figure}

As an additional consistency check relevant to the interpretation of $\mathbf{E}^{\prime} \cdot \mathbf{J}$ in Sec.~\ref{sec:Hevol}, we monitor the ratio $E_{\text{rms}}^2/B_{\text{rms}}^2$ throughout the simulations. In all cases, we find $E_{\text{rms}}^2/B_{\text{rms}}^2 \le 1$ (Fig.~\ref{fig:EvsB_app}), consistent with magnetically dominated, subluminal fluctuations~\citep{Gralla2014,Gralla2019,Narita2017}.

\section{Evaluation of the numerical value of the first kinetic Saffman integral}\label{appB}
A useful aspect of the first kinetic Saffman integral (or the ``magnetic integral'') $\mathcal{I}_{\mathscr{L}, 1} (R)$ (see, e.g., Eq.~\ref{eq:saffman2}) is that, at the reference time $t = t_0$ used to define the history term, $\mathscr{L}_1 (t_0) = h (t_0)$ (e.g., Eq.~\ref{subeq:L_1}), also equivalent to $\mathscr{L}_{\iota} (t_0)$ for initially zero bulk flows, and therefore the corresponding plateau values of the correlation integrals coincide: 
\begin{align}
I_{\mathscr{L}, 1} (t_0) = I_H (t_0)
\end{align}
and $\simeq I_{\mathscr{L} \iota} (t_0)$ for magnetically dominated systems. Under homogeneity/ergodicity and scale separation, $I_H$ corresponds to the $L \ll R \ll L_{\text{sys}}$ plateau of the volume-averaged magnetic-helicity variance. A convenient Fourier characterization is that, for statistically isotropic fields, the isotropized helicity-variance spectrum~\citep{Hosking2021,Zhou2022},
\begin{align}
\Theta (k) = \frac{S_{d - 1} k^{d - 1}}{(2\pi)^d} \int d^d r \; \langle h (\mathbf{x}) h (\mathbf{x} + \mathbf{r}) \rangle e^{-i\mathbf{k} \cdot \mathbf{r}},
\end{align}
obeys
\begin{align}
\Theta (k \to 0) = \frac{S_{d - 1}}{(2\pi)^d} I_H k^{d - 1} + o (k^{d - 1}),
\end{align}
and therefore
\begin{align}
I_H = \lim_{k \to 0} \frac{(2\pi)^d}{S_{d - 1}} k^{- (d - 1)} \Theta (k).
\end{align}
Here, $S_{d - 1} = 2\pi^{d/2}/\Gamma (d/2)$ is the surface area of a unit $(d - 1)$ sphere and $\Gamma (\cdot)$ is the gamma function. Let us now assume, in the isotropic continuum approximation, a random-phase construction (Appendix~\ref{appA}) in which the helical weights are allowed to be scale dependent:
\begin{align}
\mathbf{B}_{\mathbf{k}} = \sqrt{F(k)} \{ \sqrt{w_+ [\sigma_k (k)]} e^{i\phi_+} \hat{\mathbf{e}}_+ \nonumber \\
+ \sqrt{w_- [\sigma_k (k)]} e^{i\phi_-} \hat{\mathbf{e}}_- \}, \label{eq:Frep2}
\end{align}
where $F(k)$ sets the spectrum of the magnetic field via 
\begin{align}
B_{\text{rms}}^2 = \frac{1}{4} \frac{S_{d - 1}}{(2\pi)^d} \int_0^{\infty} dk \; k^{d - 1} F(k). \label{eq:Brms_to_Fk}
\end{align}

In Eq.~\ref{eq:Frep2}, $\mathbf{B}_{\mathbf{k}}$ is the complex synthesis amplitude used in Appendix~\ref{appA}, i.e., the coefficient that appears inside a real-field contribution $\Re \{ \mathbf{B}_{\mathbf{k}} e^{i \mathbf{k} \cdot \mathbf{x}} \}$. Expanding the real part,
\begin{align}
\Re \{ \mathbf{B}_{\mathbf{k}} e^{i \mathbf{k} \cdot \mathbf{x}} \}
= \frac{1}{2}\mathbf{B}_{\mathbf{k}} e^{i \mathbf{k} \cdot \mathbf{x}}
+ \frac{1}{2}\mathbf{B}_{\mathbf{k}}^{*} e^{-i \mathbf{k} \cdot \mathbf{x}},
\end{align}
so a single synthesis term contributes Fourier series coefficients $\boldsymbol{\mathcal{B}}_{\mathbf{k}} = \mathbf{B}_{\mathbf{k}}/2$ and $\boldsymbol{\mathcal{B}}_{-\mathbf{k}} = \mathbf{B}_{\mathbf{k}}^{*}/2$, and likewise $\boldsymbol{\mathcal{A}}_{\mathbf{k}} = \mathbf{A}_{\mathbf{k}}/2$. If both $\pm \mathbf{k}$ appear in the discrete synthesis set, then the net Fourier series coefficient at $\mathbf{k}$ is $\boldsymbol{\mathcal{B}}_{\mathbf{k}} = (\mathbf{B}_{\mathbf{k}} + \mathbf{B}_{-\mathbf{k}}^{*})/2$ and similarly for $\boldsymbol{\mathcal{A}}_{\mathbf{k}}$. For the isotropic continuum estimate below, we adopt the standard convention in which one representative per $\pm \mathbf{k}$ pair is used; with that convention, the factors $1/4$ in Eq.~\ref{eq:Brms_to_Fk2} and $1/16$ in Eq.~\ref{eq:I_H_variancedef2} follow directly from the $1/2$ factors above. With the Fourier transform convention $\tilde{\mathbf{B}} (\mathbf{k}) = \int_{V} d^d x \; \mathbf{B}(\mathbf{x}) e^{-i \mathbf{k} \cdot \mathbf{x}}$, one has $\tilde{\mathbf{B}} (\mathbf{k}) = V^{1/2} \boldsymbol{\mathcal{B}}_{\mathbf{k}}$ and similarly for $\tilde{\mathbf{A}} (\mathbf{k})$.

For each Fourier mode in Coulomb gauge, 
\begin{align}
i\mathbf{k} \times \hat{\mathbf{e}}_{\pm} = \pm k \hat{\mathbf{e}}_{\pm},
\end{align}
and thus Eq.~\ref{eq:Frep2} implies
\begin{align}
\mathbf{A}_{\mathbf{k}} = \frac{\sqrt{F (k)}}{k} \{ \sqrt{w_+ [\sigma_k (k)]} e^{i\phi_+} \hat{\mathbf{e}}_+ \nonumber \\
- \sqrt{w_- [\sigma_k (k)]} e^{i\phi_-} \hat{\mathbf{e}}_- \}. \label{eq:FrepA}
\end{align}
The Fourier transform of $h$ is the convolution
\begin{align}
\tilde{h} (\mathbf{k}) = \frac{V}{4} \int \frac{d^d k^{\prime}}{(2\pi)^d} \mathbf{A}_{\mathbf{k}^{\prime}} \cdot \mathbf{B}_{\mathbf{k} - \mathbf{k}^{\prime}}. \label{eq:hFx}
\end{align}
To connect to $\Theta (k)$, we consider, with Eq.~\ref{eq:hFx},
\begin{align}
\langle | \tilde h(\mathbf{k}) |^2 \rangle = \frac{V^2}{16} \int \frac{d^d k^{\prime}}{(2\pi)^d} \int \frac{d^d k^{\prime \prime}}{(2\pi)^d} \langle (\mathbf{A}_{\mathbf{k}^{\prime}}\cdot\mathbf{B}_{\mathbf{k}-\mathbf{k}^{\prime}}) \nonumber \\
\times (\mathbf{A}_{\mathbf{k}^{\prime \prime}}\cdot\mathbf{B}_{\mathbf{k} - \mathbf{k}^{\prime \prime}})^* \rangle.
\label{eq:hpower_start}
\end{align}
With random phases, the phase average eliminates most cross terms. However, there are two distinct phase-matching contraction channels that survive in Eq.~\ref{eq:hpower_start}:
(1) the direct contraction $\mathbf{k}^{\prime \prime} = \mathbf{k}^{\prime}$ and (2) the exchange contraction $\mathbf{k}^{\prime \prime} = \mathbf{k} - \mathbf{k}^{\prime}$. In the limit $\mathbf{k} \to 0$, these become $\mathbf{k}^{\prime \prime} = \mathbf{k}^{\prime}$ and $\mathbf{k}^{\prime \prime} \simeq - \mathbf{k}^{\prime}$, respectively, and under isotropy and $\mathbf{k} \leftrightarrow - \mathbf{k}$ symmetry the two contributions are equal. Therefore, Eq.~\ref{eq:hpower_start} gives
\begin{align}
\lim_{k\to 0} \frac{1}{V} \langle | \tilde h(\mathbf{k}) |^2 \rangle = \frac{1}{8} \int \frac{d^d k^{\prime}}{(2\pi)^d} \langle | \mathbf{A}_{\mathbf{k}^{\prime}} \cdot \mathbf{B}_{-\mathbf{k}^{\prime}} |^2 \rangle.
\label{eq:hpower_two_contractions}
\end{align}
A straightforward average over the independent phases $\phi_{\pm}$ yields, from Eqs.~\ref{eq:Frep2} and~\ref{eq:FrepA},
\begin{align}
\langle | \mathbf{A}_{\mathbf{k}^{\prime}} \cdot \mathbf{B}_{-\mathbf{k}^{\prime}} |^2 \rangle
= 2w_+ w_- \frac{F^2(k^{\prime})}{k^{\prime 2}} \nonumber \\
= \frac{1-\sigma_k^2 (k^{\prime})}{2} \frac{F^2(k^{\prime})}{k^{\prime 2}}. \label{eq:Fouriervariance}
\end{align}
Substituting Eq.~\ref{eq:Fouriervariance} into Eq.~\ref{eq:hpower_two_contractions} gives
\begin{align}
\lim_{k\to 0} \frac{1}{V} \langle |\tilde h(\mathbf{k})|^2 \rangle = \frac{1}{16} \int \frac{d^d k^{\prime}}{(2\pi)^d} [1-\sigma_k^2 (k^{\prime})] \nonumber \\
\times \frac{F^2(k^{\prime})}{k^{\prime 2}}.
\label{eq:hpower_final}
\end{align}
Combining Eq.~\ref{eq:hpower_final} with the small-$k$ form of $\Theta (k)$ above (equivalently, Eqs.~2.14 and 2.15 of~\citet{Zhou2022}) yields
\begin{align}
I_H \simeq \frac{1}{16} \frac{S_{d - 1}}{(2\pi)^d} \int_0^{\infty} dk \; [1-\sigma_k^2 (k)] k^{d - 3} F^2 (k), \label{eq:I_H_variancedef2}
\end{align}
which was referenced as Eq.~\ref{eq:I_H_variancedef}. Equations~\ref{eq:I_H_variancedef2} and~\ref{eq:Brms_to_Fk} motivate defining the effective magnetic-helicity-density correlation length
\begin{align}
L_h \equiv \bigg\{ \frac{\int_0^{\infty} dk \; k^{d - 3} F^2 (k)}{[\int_0^{\infty} dk \; k^{d - 1} F (k)]^2} \bigg\}^{1/(d + 2)}, \label{eq:emhdcl}
\end{align}
intended to characterize the scale dominating the magnetic-helicity-density two-point function $\langle h (\mathbf{x}) h (\mathbf{x} + \mathbf{r}) \rangle$.

With our specific $F (k; t = 0) = F_0 (k)$ (Eq.~\ref{eq:originalFk}) and if $\sigma_k (k; t = 0) \simeq \sigma_0$ over the contributing scales, Eq.~\ref{eq:I_H_variancedef2} becomes
\begin{align}
I_H (t = 0) \simeq (1 - \sigma_0^2) \frac{S_{d - 1}}{(2\pi)^d} A^2 e^2 k_c^{a - 1} \nonumber \\
\times 2^{-(a + 9)/2} \Gamma \bigg( \frac{a - 1}{2} \bigg). \label{eq:IHAexpression}
\end{align}
With $\langle |\mathbf{B}|^2 \rangle_{V_{\text{sys}}} = B_{\text{rms}}^2$, then, with our convention $|\mathbf{B}_{\mathbf{k}}|^2 = F(k)$, 
\begin{align}
B_{\text{rms}, 0}^2 = \frac{1}{4} \frac{S_{d - 1}}{(2\pi)^d} \int_0^{\infty} dk \; k^{d - 1} F_0 (k) = \frac{1}{8} \frac{S_{d - 1}}{(2\pi)^d} \nonumber \\
\times Ae k_c^{(a + d + 1)/2} \Gamma \bigg( \frac{a + d + 1}{4} \bigg). \label{eq:Brms_to_Fk2}
\end{align}
Substituting Eq.~\ref{eq:Brms_to_Fk2} into Eq.~\ref{eq:IHAexpression} yields
\begin{align}
I_H (t = 0) \simeq (1 - \sigma_0^2) \frac{(2\pi)^d}{S_{d - 1}} B_{\text{rms}, 0}^4 k_c^{-(d + 2)} \nonumber \\
\times 2^{-(a - 3)/2} \frac{\Gamma (\frac{a - 1}{2})}{\Gamma^2 (\frac{a + d + 1}{4})}. \label{eq:IH0estimate}
\end{align}
Equation~\ref{eq:IH0estimate} gives, in the isotropic continuum approximation, for $a = 7$, $d = 2$, $\sigma_0 = 0$, $B_{\text{rms}, 0} \; [m_e c^2/(e d_{e0})] = 0.5$, and $k_c \; [d_{e0}^{-1}] = 0.5$ (see, e.g., Table~\ref{tab:table1}), $I_H (t = 0) \; [(m_e c^2/e)^4] = 16/9$, which we use as a reference value for comparison in Figs.~\ref{fig:Saffman_evol_theory} and~\ref{fig:demo} as the black dotted lines. For the finite-$M$ thin-ring initialization used in the simulations, Eq.~\ref{eq:IH0estimate} should be interpreted as an isotropic-continuum reference estimate. The corresponding discrete expression is obtained by replacing the continuum integrals with sums over the discrete Fourier series coefficients $\boldsymbol{\mathcal{B}}_{\mathbf{k}}$ and $\boldsymbol{\mathcal{A}}_{\mathbf{k}}$ associated with the initialized real field, with $\boldsymbol{\mathcal{B}}_{\mathbf{k}}$ receiving contributions from any synthesized $\pm\mathbf{k}$ terms as described above. Access to the $k \to 0$ regime additionally requires a domain large enough to support Fourier modes well below the populated ring.

\section{On the interpretation of $B^2 L$-preserving interactions at kinetic scales}\label{appC} 
We return now to the interpretation of our 2D3V simulation results that like-helicity interactions predominantly reduce structures' magnetic helicity while leaving the interaction-scale magnetic amplitude combination $B_s^2 L_s$ only weakly altered over the event (e.g., Eq.~\ref{eq:workinghyp}), with $B_s$ and $L_s$ the structure-level magnetic field amplitude and size measures. To parametrize the helical content of a structure $s$, we introduce a structure fractional helicity $\sigma_s \equiv h_s/(B_s^2 L_s)$, where $h_s$ denotes the characteristic magnetic-helicity density in $V_s$, i.e., $h_s \sim H_{V_s}/V_s$. With this definition, combined with Eq.~\ref{eq:Hscalingeqn}, changes in $|h_s|$ decompose as 
\begin{align} \frac{d}{dt} |h_s| = B_s^2 L_s \frac{d}{dt} |\sigma_s| + |\sigma_s| \frac{d}{dt} (B_s^2 L_s) \sim L_s \frac{d}{dt} (B_s^2). \label{eq:chirality1} 
\end{align} 
Recall that in Sec.~\ref{sec:Hevol}, for globally nonhelical fields during the early kinetic stage, we found that $I_H \propto B^4$ (see, e.g., Fig.~\ref{fig:Saffman_evol_theory}) and in Sec.~\ref{sec:canonicalhelicity} we demonstrated numerical evidence that $BL \sim \text{const}$ globally (e.g., Fig.~\ref{fig:BL}). Assuming that at any given time $|\sigma_s|$ and $B_s$ are representative of the magnitude of the helical content in all magnetic structures and the global rms $B$, respectively, i.e., the system is approximately single scale, we may write the approximate scaling of the Saffman helicity-variance plateau value as 
\begin{align} 
I_H \sim h^2 L^d \sim |\sigma_s|^2 B^4 L^4 \sim |\sigma_s|^2 \propto B^4, \label{eq:C2} 
\end{align} 
and therefore, upon utilizing Eq.~\ref{eq:chirality1}, 
\begin{align} 
(B_s^2 L_s)^{-1} \frac{d}{dt} (B_s^2 L_s) \sim |\sigma_s|^{-1} (|\sigma_s|^{-1} - 1) \frac{d}{dt} |\sigma_s|. \label{eq:BsLsinter} \end{align} 
Equations~\ref{eq:C2} and~\ref{eq:BsLsinter} should be understood only as a representative-patch, rms consistency estimate. In writing $I_H \sim |\sigma_s|^2 B^4 L^4$, we have absorbed the filling factor, amplitude distribution, and sign correlation/cancellation of opposite-signed $h$ or $\mathscr{L}_1$ structures into an assumed slowly varying prefactor. Thus Eq.~\ref{eq:BsLsinter} does not by itself establish $B_s^2 L_s$ conservation for a cancellation-dominated ensemble; rather, conditional on a short event involving like-signed $\mathscr{L}_1$ structures, and assuming that the hidden ensemble prefactor and the structure geometry vary weakly during the event, Eq.~\ref{eq:BsLsinter} is compatible, particularly for $|\sigma_s| \simeq 1$, with the decrease of $|h_s|$ being carried mainly by a reduction of $|\sigma_s|$, with only subleading change in \begin{align} 
B_s^2 L_s \sim \text{const}. \label{eq:Bs2Ls} 
\end{align} 
Interactions between opposite-signed $\mathscr{L}_1$ structures represent a separate cancellation channel and enter the global Saffman second-moment argument leading to $BL \sim \text{const}$, not the local likesign coalescence interpretation. For cancellation-dominated, i.e., globally nonhelical, configurations, we return to the kinetic Saffman-style scaling derived in Sec.~\ref{sec:canonicalhelicity} (Eq.~\ref{eq:L1scaling}) to obtain the global similarity constraint $B L \sim \text{const}$ from the approximate invariance of $I_{\mathscr{L}, 1}$. 

This interaction-scale interpretation should be understood with $t_0$ chosen relative to the local interaction or short sequence of interactions, so that the accumulated source term is dominated by the event under consideration rather than by unrelated earlier evolution.

\section{Component decomposition and gauge properties of the source-compensated density}\label{appD}

\subsection{Component decomposition and fluid-limit reductions}
We decompose $\mathscr{L}_{\iota}$ into four separate source-compensated components $\mathscr{L}_{\iota, l}$, with $l \in \{ 1, 2, 3, 4 \}$, that obey their own respective continuity-type equations analogous to Eq.~\ref{eq:hLeqn1},
\begin{align}
\partial_t \mathscr{L}_{\iota, l} + \nabla \cdot \mathbf{F}_{\iota, l} = 0, \label{eq:hL_l}
\end{align}
with 
\begin{align}
(\mathscr{L}_{\iota}, \nabla \cdot \mathbf{F}_{\iota}) = \sum_{l = 1}^4 (\mathscr{L}_{\iota, l}, \nabla \cdot \mathbf{F}_{\iota, l}), \label{eq:hL_l2}
\end{align}
and it is straightforward to verify that this is satisfied with the scalar fields
\begin{subequations}
\begin{align}
\mathscr{L}_{\iota, 1} \equiv \mathscr{L}_1 \equiv h + 2c \int_{t_0}^t dt^{\prime} \; (\mathbf{E} \cdot \mathbf{B}) (t^{\prime}),
\end{align}
\begin{align}
\mathscr{L}_{\iota, 2} \equiv \frac{m_{\iota}^2 c^2}{q_{\iota}^2} \bigg[ h_{K \iota} - 2 \int_{t_0}^t dt^{\prime} \; (\pmb{\omega}_{\iota} \cdot \partial_{t^{\prime}} \mathbf{u}_{\iota}) (t^{\prime}) \bigg], \label{subeq:L_2}
\end{align}
\begin{align}
\mathscr{L}_{\iota, 3} \equiv \frac{2 m_{\iota} c}{q_{\iota}} \bigg[ h_{c \iota} + \int_{t_0}^t & dt^{\prime} \; ( \pmb{\omega}_{\iota} \cdot c \mathbf{E} \nonumber \\
& - \mathbf{B} \cdot \partial_{t^{\prime}} \mathbf{u}_{\iota} ) (t^{\prime}) \bigg], \label{subeq:L_3}
\end{align}
\begin{align}
\mathscr{L}_{\iota, 4} \equiv \frac{m_{\iota} c}{q_{\iota}} (\pmb{\omega}_{\iota} \cdot \mathbf{A} - h_{c \iota}), \label{subeq:L_4}
\end{align}
\end{subequations}
and the flux components
\begin{subequations}
\begin{align}
\mathbf{F}_{\iota, 1} \equiv \mathbf{F}_{1} \equiv - c (\mathbf{A} \times \mathbf{E} - \varphi \mathbf{B}), \label{subeq:F_1}
\end{align}
\begin{align}
\mathbf{F}_{\iota, 2} \equiv \frac{m_{\iota}^2 c^2}{q_{\iota}^2} \mathbf{u}_{\iota} \times \partial_t \mathbf{u}_{\iota}, \label{subeq:F_2}
\end{align}
\begin{align}
\mathbf{F}_{\iota, 3} \equiv - \frac{2 m_{\iota} c^2}{q_{\iota}} \mathbf{u}_{\iota} \times \mathbf{E}, \label{subeq:F_3}
\end{align}
\begin{align}
\mathbf{F}_{\iota, 4} \equiv - \frac{m_{\iota} c}{q_{\iota}} \partial_t (\mathbf{u}_{\iota} \times \mathbf{A}). \label{subeq:F_4}
\end{align}
\end{subequations} 
Here, $h_{K \iota} = \mathbf{u}_{\iota} \cdot \pmb{\omega}_{\iota}$ and $h_{c \iota} = \mathbf{u}_{\iota} \cdot \mathbf{B}$ are the kinetic- and cross-helicity densities. Note that the $l = 1$ component of Eq.~\ref{eq:hL_l} coincides with the magnetic-helicity evolution equation (Eq.~\ref{eq:Hevoleqn}). The first element of the source-compensated density (Eq.~\ref{subeq:L_1}) reduces in the ideal-MHD limit to the magnetic-helicity density, e.g., Eq.~\ref{eq:reduction_h}. In 2D MHD the second element (Eq.~\ref{subeq:L_2}) becomes the kinetic-helicity density,
\begin{align}
\frac{q_{\iota}^2 \mathscr{L}_{\iota, 2}}{m_{\iota}^2 c^2} \underset{\text{2D MHD}}{\to} h_K,
\end{align}
since $\mathbf{u} = \hat{\mathbf{z}} \times \nabla \phi$ implies $\pmb{\omega} \cdot \partial_t \mathbf{u} = 0$, with $\phi$ the stream function. In incompressible ideal MHD with $\rho = \text{const}$, the third element (Eq.~\ref{subeq:L_3}) reduces to the cross-helicity density,
\begin{align}
\frac{q_{\iota} \mathscr{L}_{\iota, 3}}{2 m_{\iota} c} \underset{\text{inc-ideal MHD}}{\to} h_c,
\end{align}
because $\pmb{\omega} \cdot c \mathbf{E} - \mathbf{B} \cdot \partial_t \mathbf{u}$ becomes a pure divergence that can be absorbed into $\mathbf{F}_{\iota, 3}$. The fourth element (Eq.~\ref{subeq:L_4}) shows that $\pmb{\omega}_{\iota} \cdot \mathbf{A}$ differs from the cross-helicity density up to a divergence.

\subsection{Gauge properties}
It is worthwhile to check whether $\mathscr{L}_{\iota}$ and $\mathscr{L}_{\iota, l}$ are gauge invariant under the transformations $\mathbf{A} \to \mathbf{A}^{\prime} = \mathbf{A} + \nabla \chi$ and $c \varphi \to c \varphi^{\prime} = c \varphi - \partial_t \chi$. For the magnetic component, the time-integrated source term
$2c\int_{t_0}^t dt^{\prime} \; \mathbf{E} \cdot \mathbf{B}$ is gauge invariant, so
$\delta_{\chi} \mathscr{L}_1 = \delta_{\chi} (\mathbf{A} \cdot \mathbf{B})
= (\nabla \chi) \cdot \mathbf{B}
= \nabla \cdot (\chi \mathbf{B})$, where $\nabla \cdot \mathbf{B} = 0$. One obtains
\begin{subequations}
\begin{align}
\int_V d^d x \; \delta_{\chi} \mathscr{L}_{\iota} = \oint_{\partial V} \chi \pmb{\Omega}_{\iota} \cdot d\mathbf{S}, \label{subeq:Liota_gaugeinvariance}
\end{align}
\begin{align}
\int_V d^d x \; \delta_{\chi} \mathscr{L}_{1} = \oint_{\partial V} \chi \mathbf{B} \cdot d\mathbf{S},
\end{align}
\begin{align}
\int_V d^d x \; \delta_{\chi} \mathscr{L}_{\iota, 2} = \int_V d^d x \; \delta_{\chi} \mathscr{L}_{\iota, 3} = 0,
\end{align}
\begin{align}
\int_V d^d x \; \delta_{\chi} \mathscr{L}_{\iota, 4} = \frac{m_{\iota} c}{q_{\iota}} \oint_{\partial V} \chi \pmb{\omega}_{\iota} \cdot d\mathbf{S}.
\end{align}
\end{subequations} 
Hence, $\mathscr{L}_{1}$ (respectively, $\mathscr{L}_{\iota, 4}$) is gauge invariant when $\mathbf{B} \cdot \hat{\mathbf{n}} = 0$ (respectively, $\pmb{\omega}_{\iota} \cdot \hat{\mathbf{n}} = 0$) on $\partial V$, while $\mathscr{L}_{\iota, 2}$ and $\mathscr{L}_{\iota, 3}$ are gauge invariant without additional boundary constraints. From Eq.~\ref{subeq:Liota_gaugeinvariance}, gauge invariance of $\mathscr{L}_{\iota}$ is guaranteed if $\pmb{\Omega}_{\iota} \cdot \hat{\mathbf{n}} = 0$ on $\partial V$, which can be achieved if we assume that canonical vorticity structures are localized and arbitrarily small compared to $V$ as $V \to \infty$. This is the analogous argument to that of \citet{Hosking2021}, applied in our case to the canonical vorticity instead of the magnetic field.


\bibliography{main_v4}

\end{document}